\documentclass{article}
\usepackage{bm}
\usepackage{amsfonts, amsmath, amssymb}
\usepackage{placeins} % for \FloatBarrier
\usepackage[margin=1in]{geometry}
\usepackage[affil-it]{authblk} % for information of Department et al.
\usepackage{stmaryrd} % to use \llbracket
\usepackage{setspace} % line space for bib
\usepackage{floatrow} % for figures
\usepackage{subfig} % for figures
\usepackage[export]{adjustbox} % for two figures alignment
\usepackage{sidecap} % for figures
\usepackage{graphicx}
\usepackage[labelsep=period]{caption} % for the caption, we don't want colon
\usepackage{xcolor}
\usepackage{cite} % multiple references
\usepackage{times} % font change to times new roman
\usepackage{mathptmx} % font change to times new roman
\newcommand\keywords[1]%
  {\begin{flushleft}
   \let\and\\%
   \textbf{Keywords:}
   #1
   \end{flushleft}%
  }
  
\begin{document}
\title{Control of a sedimenting elliptical particle by electromagnetic forces}
\author[1,2,3]{Jianhua Qin}
\author[1,2,*]{Guodan Dong}
\author[3,**]{Hui Zhang}
\renewcommand\Affilfont{\fontsize{9}{10.8}\itshape}
\affil[1]{The State Key Laboratory of Nonlinear Mechanics, Institute of Mechanics, Chinese Academy of Sciences, \protect\\Beijing 100190, China.}
\affil[2]{University of Chinese Academy of Sciences, Beijing 100049, China.}
\affil[3]{Key Laboratory of Transit Physics, 
Nanjing University of Science  and Technology, Nanjing, Jiangsu 210094, China}
\affil[*]{Corresponding author at: The State Key Laboratory of Nonlinear Mechanics, Institute of Mechanics, \protect\\Chinese Academy of Sciences, Beijing 100190, China. \protect\\
University of Chinese Academy of Sciences, Beijing 100049, China.\protect\\Email: Dongguodan9@163.com}
\affil[**]{Corresponding author at: Key Laboratory of Transit Physics, Nanjing University of Science  and Technology, \protect\\Nanjing, Jiangsu 210094, China\protect\\Email: zhanghui1902@hotmail.com}
%xyang@imech.ac.cn}
\date{}

\maketitle
\section{Abstract}
In this paper, the effectiveness of electromagnetic forces on controlling the motion of a sedimenting elliptical particle is investigated using the immersed interface-lattice Boltzmann method (II-LBM), in which a signed distance function is adopted to apply the jump conditions for the II-LBM and to add external electromagnetic forces. First, mechanisms of electromagnetic control on suppressing vorticity generation based on the vorticity equation and vortex shedding based on the streamwise momentum equation are discussed. Then, systematical investigations are performed to quantify and qualify the effects of the electromagnetic control by changing the electromagnetic strength, the initial orientation angle of the elliptical particle, and the density ratio of the particle to the fluid. To demonstrate the control effect of different cases, comparisons of vorticity fields, particle trajectories, orientation angles, and energy transfers of the particles are presented. Results show that the rotational motion of the particle can be well controlled by appropriate magnitudes of electromagnetic forces. \textcolor{black}{In a relatively high solid to fluid density ratio case where vortex shedding appears, the sedimentation speed can increase nearly 40\% and the motion of the particle turns into a steady descent motion once an appropriate magnitude of the electromagnetic force is applied. When the magnitude of the electromagnetic force is excessive, the particle will deviate from the center of the side walls. In addition, the controlling approach is shown to be robust for various initial orientation angles and solid to fluid density ratios.}

\keywords{rotational control, electromagnetic control, energy transfer, immersed interface-lattice Boltzmann method.}
\section{Introduction}
For flow around a blunt body, when the Reynolds number (the ratio of inertial forces to viscous forces) grows, the flow will be transformed from a steady state to an unsteady state and vortex shedding appears. Vortex shedding is beneficial to harvest fluidic energies using piezoelectric materials. However, in most cases, the shedding \textcolor{black}{vortices} will bring bad effects like additional drag force for a travelling ship, aerodynamic noise for an airplane, undesired rotation of a missile, and structural damage for a bridge~etc. Hence, it is imperative to study the suppression of vortex shedding to lower the damage. In real life, the motions of airplanes, submarines and parachutes all involve multiple degrees of freedom, including the translational motion and the rotational motion, in which the rotational motion exists when vortex shedding appears. Specifically, freely falling or rising of bodies are very common phenomena occurred in situations like falling leaves, floating bubbles and ball games etc. When a body sediments or rises in the fluid, many possible motions can happen if the motion of the body is not controlled, including descending along a straight line, fluttering, and tumbling~\cite{ern2012wake,andersen2005unsteady,gazzola2011simulations}.  Therefore, it is meaningful to investigate the rotational control of a sedimenting body. Traditionally, the control of the rotational motion of a body is relied on imposing additional torques~\cite{chaturvedi2011rigid,udwadia2012unified}. In this work, a novel idea of controlling the rotational motion of a rigid body in the fluid by the electromagnetic force is proposed and validated through numerical simulations.

The electromagnetic control can be originated from the guess of Gialitis and Lielausis~\cite{gailitis1961possibility}. They put forward the idea that the electromagnetic force can change the structure of the boundary layer. The electromagnetic force, i.e. the Lorentz force, is generated from staggered electrodes and magnets and is parallel to the flow direction~\cite{tsinober1990mhd, crawford1997reynolds,weier1998experiments,weier2004control}. The magnitude of the electromagnetic force decreases exponentially as the penetration depth of the electromagnetic force increases. To optimize the control efficiency, Weier and Gerbeth~\cite{weier2004control} compared the control results of periodic electromagnetic forces, static electromagnetic forces, and the traditional blowing control approach. Chen and Aubry~\cite{chen2005active} proposed to produce electromagnetic forces related to the azimuthal angle outside the cylinder, and validated that the approach is able to suppress vortex shedding and reduce the drag force. Dousset and Poth{\'e}rat~\cite{dousset2008numerical} investigated flow structures for various Reynolds numbers and Hartmann numbers (the ratio of electromagnetic forces to viscous forces) under the effect of axial electromagnetic forces. Singhah and Sinhamahapatra~\cite{singha2011control} and Chatterjee~et~al.~\cite{chatterjee2012control} investigated the influence of the Hartmann number for the control effect of flow around a blunt body. Zhang~et~al.~\cite{zhang2010effect,zhang2014numerical,zhang2017numerical} numerically studied the control of wake structures of a static circular cylinder or a circular cylinder undergoing translational motions by different ways of imposing electromagnetic forces. \textcolor{black}{Fisher~et~al.~\cite{Fisher2018} presented the control of the hydraulic jump in a liquid channel flow with electromagnetic forces by experiments.} The above mentioned electromagnetic flow control approaches have only been applied to control  "constrained" boundaries such as static bodies or vortex-induced vibration of a spring mounted cylinder undergoing translational motions. To the best of our knowledge, research about the effectiveness of electromagnetic forces on controlling a freely moving body in the fluid that involves the rotational motion is presently lacking.

Due to the development of numerical methods and the improvement of experimental conditions, meticulous investigations about bodies freely rising or falling in fluids become popular. On the numerical simulation domain, to avoid time-consurming mesh reconstruction and mesh distortion, the non-body fitted mesh methods are superior in simulating large amplitude motions like falling or rising of a body in the fluid. The immersed boundary method is a representative of these non-body fitted mesh methods~\cite{peskin2002immersed,sotiropoulos2014immersed,griffith2020immersed}.
Compared to the classical immersed boundary (IB) method~\cite{peskin2002immersed}, the direct forcing IB method~\cite{uhlmann2005immersed} has an advantage in simulating rigid body dynamics because larger time steps can be used. As mentioned before, the electromagnetic force is related to the penetration depth into the fluid. This requires the calculation of the distance between a fluid point and its orthogonal projection on the body surface. However, the standard direct forcing IB method does not involve the solution to distance between a fluid point and the interface. The immersed interface-lattice Boltzmann method (II-LBM) proposed by Qin~et~al.~\cite{QIN2020109807} uses a signed distance function to track the interface so that the distance between a fluid point and its orthogonal projection on the interface can be calculated. The II-LBM is better than the immersed boundary-lattice Boltzmann method in volume conservation and higher order of accuracy. For these reasons, this study uses the II-LBM to study the control of the rotational motion of a sedimenting body via the azimuthal electromagnetic force~\cite{chen2005active}. 

Particles of nonspherical shapes are commonly encountered in the nature~\cite{subramaniam2005non}, the industry~\cite{udoh2017nanocomposite} and scientific studies~\cite{li2012sphericities,walayat2020sedimentation,huang2017ellipsoidal}. Many interesting phenomena can appear when the fluid meets nonspherical particles. Feng et~al.~\cite{feng1994direct} reported that when settling between two parallel walls, an elliptical particle will translate itself to the center of walls because of wall effects and orient its major axis to be orthogonal to gravity direction with various initial configurations. Huang~et~al.~\cite{huang_hu_joseph_1998} showed that a settling ellipse will turn its major axis to the vertical direction for sufficiently small Reynolds numbers. Xia~et~al.~\cite{xia2009flow} presented that fascinating modes including oscillating, tumbling, horizontal, vertical, and inclined sedimention could appear for an elliptical particle settling in narrow channels. \textcolor{black}{By studyding the sedimentation of an ellipsoidal particle in circular and squares tubes, Huang et~al.~\cite{Huang2014} showed that the geometry of the tube and the moment of inertia of the particle are significant to the sedimentation mode.} Zhao~et~al.~\cite{zhao2015rotation} demonstrated that preferential orientations of the nonspherical particles affect the rotation of aspherical particles. As a first step towards understanding the controlling effectiveness, this paper studies the electromagnetic control of an elliptical cylinder sedimenting in a relatively wide channel to avoid much more complicated phenomena that may appear for narrow channels~\cite{xia2009flow}. Particularly, our study is based on the sedimentation of an elliptical particle studied by Xia~et al.~\cite{xia2009flow}, in which the aspect ratio of the channel width to the major axis is equal to four.

\section{Numerical method}
In this study, we use the immersed interface-lattice Boltzmann method (II-LBM)~\cite{QIN2020109807} to simulate the sedimentation of an elliptical particle in the incompressible Newtonian fluid, and the azimuthal electromagnetic force to control the sedimenting process. This section introduces the numerical implementations of the II-LBM and the azimuthal electromagnetic force.

\subsection{The immersed interface-lattice Boltzmann method}
\subsubsection{Governing equations}
The lattice Boltzmann equations for the incompressible fluid flow using the multi-relaxation-time (MRT) operators can be written as~\cite{QIN2020109807}
\begin{equation}
\frac{\partial f_i}{\partial t}+\textit{\textbf{e}}_i\cdot \nabla f_i=-\bm{M}^{-1}\bm{S}_d\left(m_i(\bm{x},t)-m_i^{(eq)}(\bm{x},t)\right)+
\frac{1}{c_\text{s}^2}\omega_i\bm{e}_i\cdot \bm{g}(\bm{x},t)+
\frac{1}{c_\text{s}^2}\omega_i\bm{e}_i\cdot \bm{g}_\text{m}(\bm{x},t),
\label{IB_LBM_continuous}
\end{equation}
in which $f_i$ are the distribution functions, $\bm{e}_i$ are directional velocity vectors. The first term on the right hand side (RHS) of Eq.~(\ref{IB_LBM_continuous}) deals with the local collision process, in which $\bm{M}$ is the transformation matrix for the multi-relaxation-time (MRT) model~\cite{lallemand2000theory}, $\bm{S}_d$ is the diagonal matrix. The space of distribution functions ($\bm{f}=\{f_i\}$) and the space of the moments of the distribution functions ($\bm{m}=\{m_i\}$) are related by $\bm{m}=\bm{M}\bm{f}$. Similarly, we have $\bm{m}^{(eq)}=\bm{M}\bm{f}^{(eq)}$, in which $\bm{f}^{(eq)}$ and $\bm{m}^{(eq)}$ are spaces of the equilibrium distribution functions and moments of the equilibrium distribution functions, respectively. The second term on the RHS of Eq.~(\ref{IB_LBM_continuous}) includes the Eulerian force density $\bm{g}$ representing the additional body forces imposed on the fluid by the immersed boundary~\cite{he1997lattice}. The last term on the RHS of Eq.~(\ref{IB_LBM_continuous}) accounts for the effect of the electromagnetic force on the fluid, in which $\bm{g}_\text{m}$ is the electromagnetic force density.

The interactions between the fluid and the immersed boundary are formulated by
\begin{equation}
\bm{g}(\textbf{\textit{x}},t)=
\int_{\Gamma}\bm{G}(l,t)\,\delta(\textbf{\textit{x}}-\textbf{\textit{X}})
\,\text{d}l,
\label{eq:Lag_Eul_con_force}
\end{equation}
\begin{equation}
\bm{G}(l,t)=\mathcal{G}(\bm{X},\bm{U},l,t),
\label{force_direct}
\end{equation}
\begin{equation}
\frac{\partial \textbf{\textit{X}}(l,t)}{\partial t}=
\textbf{\textit{U}}(l,t)=
\int_{\Omega}\textit{\textbf{u}}(\textbf{\textit{x}},t)\,\delta (\textbf{\textit{x}}-\textbf{\textit{X}}(l,t))\,\text{d}\textbf{\textit{x}},
\label{eq:Lag_pos_update_con}
\end{equation}
in which $\Gamma$ and $\Omega$ are the immersed boundary and the fluid domain, respectively, \textcolor{black}{$l$ is the curvilinear coordinate that parameterizes the interface,} $\bm{G}(l,t)$ is the Lagrangian force density, $\bm{X}(l,t)$ and $\bm{U}(l,t)$ are the position and velocity of the boundary, and $\mathcal{G}$ is the functional that determines the interfacial force from the deformations and/or velocities of the immersed boundary. $\delta (\bm{x}-\bm{X}(l,t))$ is the Dirac delta function used to spread the Lagrangian forces to the Eulerian force and integrate the Eulerian velocity to obtain the Lagrangian velocity.

The difference between the II-LBM and the IB-LBM is that the former formulation considers the jump conditions of the distribution functions so that higher order of accuracy and substantial better volume conservation can be achieved~\cite{QIN2020109807}. The jump conditions can be written as
\begin{equation}
\llbracket f_i\rrbracket=\frac{\omega_i}{c_\text{s}^2}
\frac{\bm{G}(l,t)\cdot \bm{n}}{|\frac{\text{d} \bm{X}}{\text{d} l}|},
\label{eq:jump_condition}
\end{equation}
in which $\omega_i$ is the weighting function in each direction, $\bm{n}$ is the unit normal vector to the interface, $c_\text{s}$ is the sound speed in lattice units, and $|\frac{\text{d} \bm{X}}{\text{d} l}|=1$ for rigid bodies.

\subsubsection{Numerical discretizations}
The lattice Boltzmann equation (LBE) is discretized via finite differences in both time and space, resulting in the following discretized LBE
\begin{equation}
\begin{aligned}
& \frac{f_i(\bm{x},t+\Delta t)-f_i(\bm{x},t)}{\Delta t}+
\frac{f_i(\bm{x},t)-
f_i(\bm{x}-\Delta \bm{x}_i,t)-\llbracket f_i(\bm{x},t)\rrbracket}{\Delta t}\\
=& -\textit{\textbf{M}}^{-1}\textit{\textbf{S}}_d\left(m_i(\bm{x}-\Delta \bm{x}_i,t)-m_i^{(\text{eq})}(\bm{x}-\Delta \bm{x}_i,t)\right)+\frac{1}{c_\text{s}^2}\omega_i\bm{e}_i\cdot
\bm{g}^{||}(\bm{x},t)+
\frac{1}{c_\text{s}^2}\omega_i\bm{e}_i\cdot \bm{g}_\text{m}(\bm{x},t),
\end{aligned}
\label{eq:discretized_LBE_He}
\end{equation}
in which $\bm{g}^{||}(\bm{x},t)$ is the tangential part of the Eulerian force, and $\Delta \bm{x}_i=\bm{e}_i\Delta t$ is the grid spacing in the $\bm{e}_i$ direction. Here, we use the jump conditions to deal with the normal part of the Eulerian force~\cite{QIN2020109807},

For the MRT D2Q9 model (i.e.~$i=0,1,2,...,8$), the weighting functions are defined via
\begin{equation}
\omega_i=
\begin{cases}
\frac{4}{9},&i=0,\\
\frac{1}{9},&1\leq i \leq4,\\
\frac{1}{36},&\text{otherwise},
\end{cases}
\label{eq:omegai}
\end{equation}
and the directional velocity vectors $\bm{e}_i$ are defined as $(0,0)$, $(\pm1,0)$, $(0,\pm1)$, $(\pm1,\pm1)$. Then the macroscopic density, pressure and velocity can be obtained via $\rho_\text{f}=\sum_if_i$, $p=\frac{1}{3}\rho_\text{f}$ and $\bm{u}=\frac{1}{\rho_\text{f}}\sum_if_i\bm{e}_i$. The transformation matrix and diagonal matrix for the D2Q9 model are
\begin{equation}
\bm{M}=
\setcounter{MaxMatrixCols}{10}
\small{
\begin{bmatrix}
1 & 1 & 1 & 1 & 1 & 1 & 1 & 1 & 1 \\
-4 & -1 & -1 & -1 & -1 & 2 & 2 & 2 & 2 \\
4 & -2 & -2 & -2 & -2 & 1 & 1 & 1 & 1 \\
0 & 1 & 0 & -1 & 0 & 1 & -1 & -1 & 1 \\
0 & -2 & 0 & 2 & 0 & 1 & -1 & -1 & 1 \\
0 & 0 & 1 & 0 & -1 & 1 & 1 & -1 & -1 \\
0 & 0 & -2 & 0 & 2 & 1 & 1 & -1 & -1 \\
0 & 1 & -1 & 1 & -1 & 0 & 0 & 0 & 0 \\
0 & 0 & 0 & 0 & 0 & 1 & -1 & 1 & -1 \\
\end{bmatrix},
}
\end{equation}
and $\bm{S}_\text{d}=\text{diag}(s_0,s_1,...,s_8)$, in which $s_0=s_3=s_5$, $s_1=\frac{1.63}{\Delta t}$, $s_2=\frac{1.14}{\Delta t}$, $s_4=s_6=\frac{1.92}{\Delta t}$, and $s_7=s_8=\frac{1}{\lambda}$. When solving the equations, we use $\Delta x=\Delta t=1$ for the space and time steps. The force on the $k^\text{th}$ Lagrangian boundary point can be obtained via a direct-focing IB approach as
\begin{equation}
\bm{G}_k(t)=\rho_\text{f}\frac{\bm{U}^{\text{D}}_k(t)-\bm{U}^*_k(t)}{\Delta t},
\end{equation}
where $\bm{U}^{\text{D}}_k(t)$ is the desired Lagrangian velocity, and $\bm{U}^*_k(t)$ is the intermediate Lagrangian velocity obtained by
\begin{equation}
\bm{U}^*_k(t) = \sum_{\bm{x}}\bm{u}^*(\bm{x},t)\delta_h(\bm{x}-\bm{X}_k(t))\Delta x^2,
\end{equation}
where $\bm{u}^*(\bm{x},t)$ is the fluid velocity obtained by solving the fluid flow equations without considering the immersed boundary, and $\bm{X}_k$ is the Lagrangian coordinate. Here, $\delta_h(\bm{x}-\bm{X})$ is a regularized delta function, which we define by
\begin{equation}
\delta_h (\bm{x}-\bm{X})=
\frac{1}{\left(\Delta x\right)^3}\,\Phi\! \left(\frac{x-X}{\Delta x}\right)\,
\Phi\!\left(\frac{y-Y}{\Delta x}\right)\,
\Phi\!\left(\frac{z-Z}{\Delta x}\right),
\end{equation}
in which $\Phi(r)$ is a one-dimensional kernel function. We use Peskin's four-point kernel function~\cite{peskin2002immersed},
\begin{equation}
\Phi (r)=
\begin{cases}
\frac{1}{8}\left(3-2|r|+\sqrt{1+4|r|-4r^2}\right), & |r|\leq 1,\\
\frac{1}{8}\left(5-2|r|-\sqrt{-7+12|r|-4r^2}\right), & 1\leq |r|\leq 2,\\ 
0,& |r|>2 .\\
\end{cases}
\end{equation}The Eulerian force is obtained from the distribution of the Lagrangian force via the same $\delta_h$
\begin{equation}
\bm{g}(\bm{x},t)=\sum_k\bm{G}_k(t)\delta_h(\bm{x}-\bm{X}_k(t))\Delta X,
\end{equation} 
in which $\Delta X$ is the distance between two adjacent Lagrangian points. In this investigation, $\Delta X=\Delta x$ is chosen. Then $\bm{g}^{||}(\bm{x},t)$ can be calculated via
\begin{equation}
\bm{g}^{||}(\bm{x},t)=
\sum_k\bm{G}^{||}_k(t)\,\delta_h
(\bm{x}-\bm{X}_k(t))\,\Delta X,
\label{Eq_IBM_tangential}
\end{equation}
in which $\bm{G}^{||}_k=\bm{G}_k-
(\bm{G}_k\cdot\bm{n})\bm{n}$ is the tangential part of the Lagrangian force, and $\bm{n}$ is the outward normal direction with respect to the interface. Then the only unknown in the discretized II-LBM equation is $\llbracket f_i\rrbracket$. To calculate the jump conditions, $\bm{n}$ and $\bm{G}(l,t)$ in Eq.~(\ref{eq:jump_condition}) have to be determined. The calculations of the two variables will be introduced along with the identification of the interface later in Sec.~\ref{sec:interface_repres}. 

%In this study, the translational and rotational motions of an elliptical particle are simulated. The equations for the motion of the elliptical particle are
%\begin{equation}
%m\bm{a}=-\int_\Gamma \bm{G}(l,t)\text{d}l+(\rho_\text{s}-\rho_\text{f})V_\text{s}\bm{g}_\text{e}+\rho_\text{f}V_s\frac{\text{d}\bm{U}}{\text{d}t},
%\label{eq:acce_continuous}
%\end{equation}
%\begin{equation}
%\mathbb{I}_\text{s}\bm{w} = -\int_\Gamma (\bm{X}(l,t)-\bm{X}_\text{c})\times \bm{G}(l,t)\text{d}l+\frac{\rho_\text{f}}{\rho_\text{s}}\mathbb{I}_\text{s}\frac{\text{d}\bm{W}}{\text{d}t},
%\label{eq:ang_acce_continuous}
%\end{equation}
%in which $\bm{a}=\frac{\text{d}\bm{U}_\text{c}}{\text{d}t}$, $\bm{w}=\frac{\text{d}\bm{W}}{\text{d}t}$, and $\bm{W}$ are linear acceleration, angular acceleration and angular velocity of the rigid body, respectively. The first terms on the right hand side of Eqs.~(\ref{eq:acce_continuous}) and (\ref{eq:ang_acce_continuous}) represent the force and torque exerted on the rigid body by the ambient fluid, respectively. The last terms in Eqs.~(\ref{eq:acce_continuous}) and (\ref{eq:ang_acce_continuous}) account for the motion of the fluid inside the rigid body~\cite{feng2009robust}.

In this study, the translational and rotational motions of an elliptical particle are simulated. The discretized equations for the motion of the elliptical particle are
\begin{equation}
m\bm{a}(t+\Delta t)=-\sum_{\bm{X}_k} \bm{G}_k\Delta X+(\rho_\text{s}-\rho_\text{f})V_\text{s}\bm{g}_\text{e}+\rho_\text{f}V_s\frac{\bm{U}(t)-\bm{U}(t-\Delta t)}{\Delta t},
\label{eq:acce_continuous}
\end{equation}
\begin{equation}
\mathbb{I}\bm{w}(t+\Delta t) = -\sum_{\bm{X}_k}(\bm{X}_k-\bm{X}_\text{c})\times \bm{G}_k\Delta X+\frac{\rho_\text{f}}{\rho_\text{s}}\mathbb{I}\frac{\bm{W}(t)-\bm{W}(t-\Delta t)}{\text{d}t},
\label{eq:ang_acce_continuous}
\end{equation}
in which $\mathbb{I}$, $\bm{a}$, $\bm{w}$, and $\bm{W}$ are moment of inertia, linear acceleration, angular acceleration and angular velocity of the rigid body, respectively. The first terms on the RHS of Eqs.~(\ref{eq:acce_continuous}) and (\ref{eq:ang_acce_continuous}) represent the force and torque exerted on the rigid body by the ambient fluid, respectively. The last terms in Eqs.~(\ref{eq:acce_continuous}) and (\ref{eq:ang_acce_continuous}) account for the motion of the fluid inside the rigid body~\cite{feng2009robust}. Then the velocity of the mass center $\bm{U}_c$ and angular velocity of the body $\bm{W}$ are updated via
\begin{equation}
\bm{U}_{\text{c}}(t+\Delta t)=\bm{U}_{\text{c}}(t)+\Delta t\,\bm{a}(t+\Delta t),
\label{eq:vel}
\end{equation}
\begin{equation}
\bm{W}(t+\Delta t)=\bm{W}(t)+\Delta t\, \bm{w}(t+\Delta t).
\label{eq:angular}
\end{equation}
Because the numerical simulations performed in this paper are in two spatial dimensions, we only need to track the orientation angle in the $z$ direction and the position of the center of mass of the rigid body. A forward Euler method is used to calculate the position of the center of mass and the orientation angle of the body
\begin{equation}
\bm{X}_{\text{c}}(t+\Delta t)=\bm{X}_{\text{c}}(t)+\bm{U}_{\text{c}}(t)\Delta t+\frac{1}{2}(\Delta t)^2\bm{a}(t+\Delta t),
\label{eq:3dof_pos}
\end{equation}
\begin{equation}
\theta(t+\Delta t)=\theta(t)+W_z(t)\Delta t+\frac{1}{2}(\Delta t)^2w_z(t+\Delta t),
\label{eq:3dof_angle}
\end{equation}
in which $\theta$, $W_z$, and $w_z$ are the orientation angle, $z$ component of the angular velocity, and the angular acceleration of the rigid body, respectively.

\subsubsection{Interface representation}
\label{sec:interface_repres}
The signed distance function is used to represent the fluid-structure interface~\cite{li2006immersed,QIN2020109807,zhu2021variationally} and calculate $\bm{n}$ and $\bm{G}(l,t)$ in Eq.~(\ref{eq:jump_condition}). Here, a brief introduction of the numerical implementation is given, the detailed implementation can be referred to Li and Ito~\cite{li2006immersed} and Qin et~al.~\cite{QIN2020109807}. For an elliptical particle centered at $(x_0,y_0)$ and immersed in a background fluid, a signed distance function $\varphi (\bm{x})$ can be defined as
\begin{equation}
\varphi(\bm{x})=\sqrt{\frac{[(x-x_0)\cos\theta+(y-y_0)\sin\theta]^2}{a^2}+\frac{[(x-x_0)\sin\theta-(y-y_0)\cos\theta]^2}{b^2}}-1,
\end{equation}
in which $a$ and $b$ are semi-major and semi-minor axes, respectively. Here, $\theta$ is the orientation angle of the elliptical particle. With the definition of $\varphi (\bm{x})$, the boundary of the particle, the exterior and interior fluid regions are represented by $\varphi (\bm{x})=0$, $\varphi (\bm{x})>0$ and $\varphi (\bm{x})<0$, respectively. Using the signed distance function, the outward unit normal vector with respect to the interface can be obtained via
\begin{equation}
\bm{n}(\bm{x})=\frac{\nabla \varphi}{|\nabla \varphi|},
\label{eq:normal_in_jump}
\end{equation}
where $\nabla \varphi(\bm{x})=(\varphi_x (\bm{x}),\varphi_y (\bm{x}))$ is approximated using second order centered differences.

If $\phi (\bm{x})\phi (\bm{x}-\Delta\bm{x}_i)<0$, $\llbracket f_i\rrbracket$ needs to be evaluated and implemented in Eq.~(\ref{eq:discretized_LBE_He}). Otherwise, $\llbracket f_i\rrbracket$ should be zero. The jump conditions for the distribution functions are accounted for through correction terms that are evaluated on the Eulerian grid points. Consequently, the boundary point $\bm{X}^*$ corresponds to a particular Eulerian point $\bm{x}$ must be determined in order to obtain $\bm{G}(l,t)$ in Eq.~(\ref{eq:jump_condition}). By using the same signed distance function, the orthogonal projection of a point $\bm{x}$ close to the interface is
\begin{equation}
\bm{X}^*=\bm{x}+r_\text{w} \nabla\varphi (\bm{x}),
\label{eq:projection_point}
\end{equation}
in which $r_\text{w}$ is the distance between $\bm{x}$ and $\bm{X}^*$ and is obtained by solving a quadratic equation obtained from the truncated Taylor expansion~\cite{QIN2020109807}. Then $\bm{G}(l,t)$ on $\bm{X}^*$ can be calculated by using the direct forcing IB method~\cite{qin2020efficient,jianhua2018numerical,QIN2020109807}.

\subsection{The azimuthal electromagnetic force}
\label{sec:magnetic}
\begin{figure}[htb!]
\centering
\includegraphics[width=8cm]{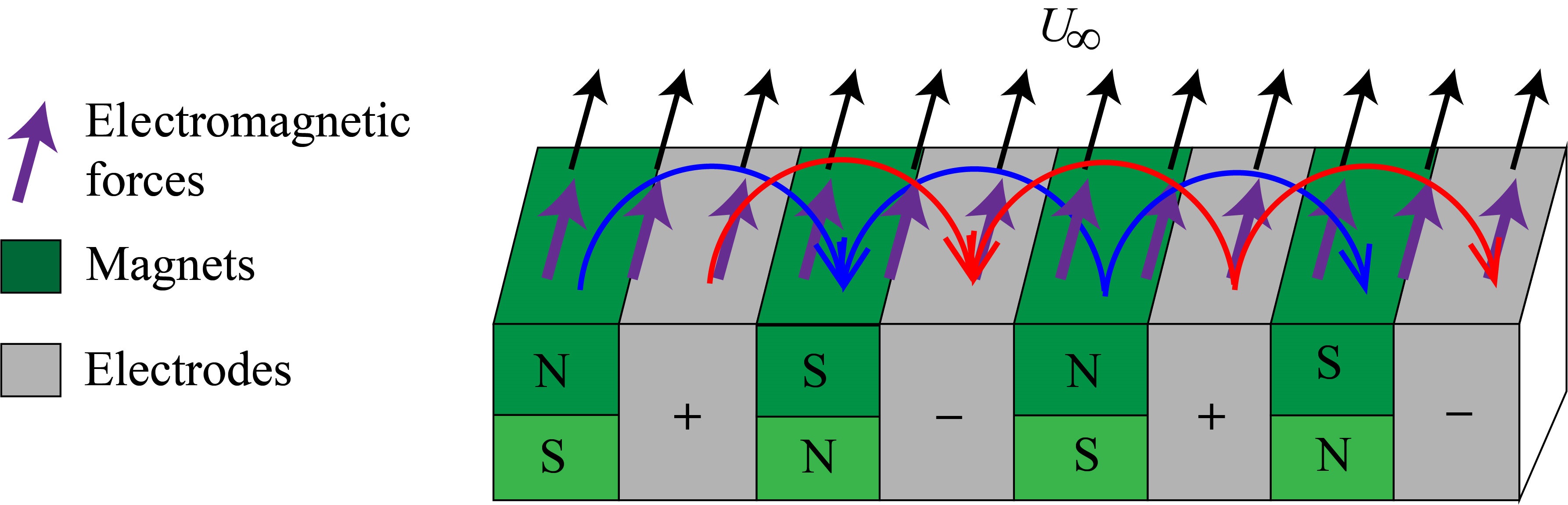}
\caption{Schematic of electromagnetic forces generated from staggered electrodes and magnets.}
\label{fig:Lorenz_Weier}
\end{figure}

In our present study, the azimuthal electromagnetic force proposed by Chen and Aubry~\cite{chen2005active} is applied on a certain range of fluids exterior to the rigid body. As shown in Fig.~\ref{fig:Lorenz_Weier}, the electromagnetic force is generated from staggered electrodes and magnets and is parallel to the flow direction~\cite{weier2004control}. For rigid body simulations, a big difference of the II-LBM compared to conventional IB-LBMs is that the exterior and interior fluid regions of the interface are described by a signed distance function for the II-LBM. Therefore, the electromagnetic force is taken into account by using the same singed distance function used in the II-LBM.

%\FloatBarrier

\begin{figure}[htb!]
\subfloat[]{\includegraphics[width=5.2cm,valign=m]{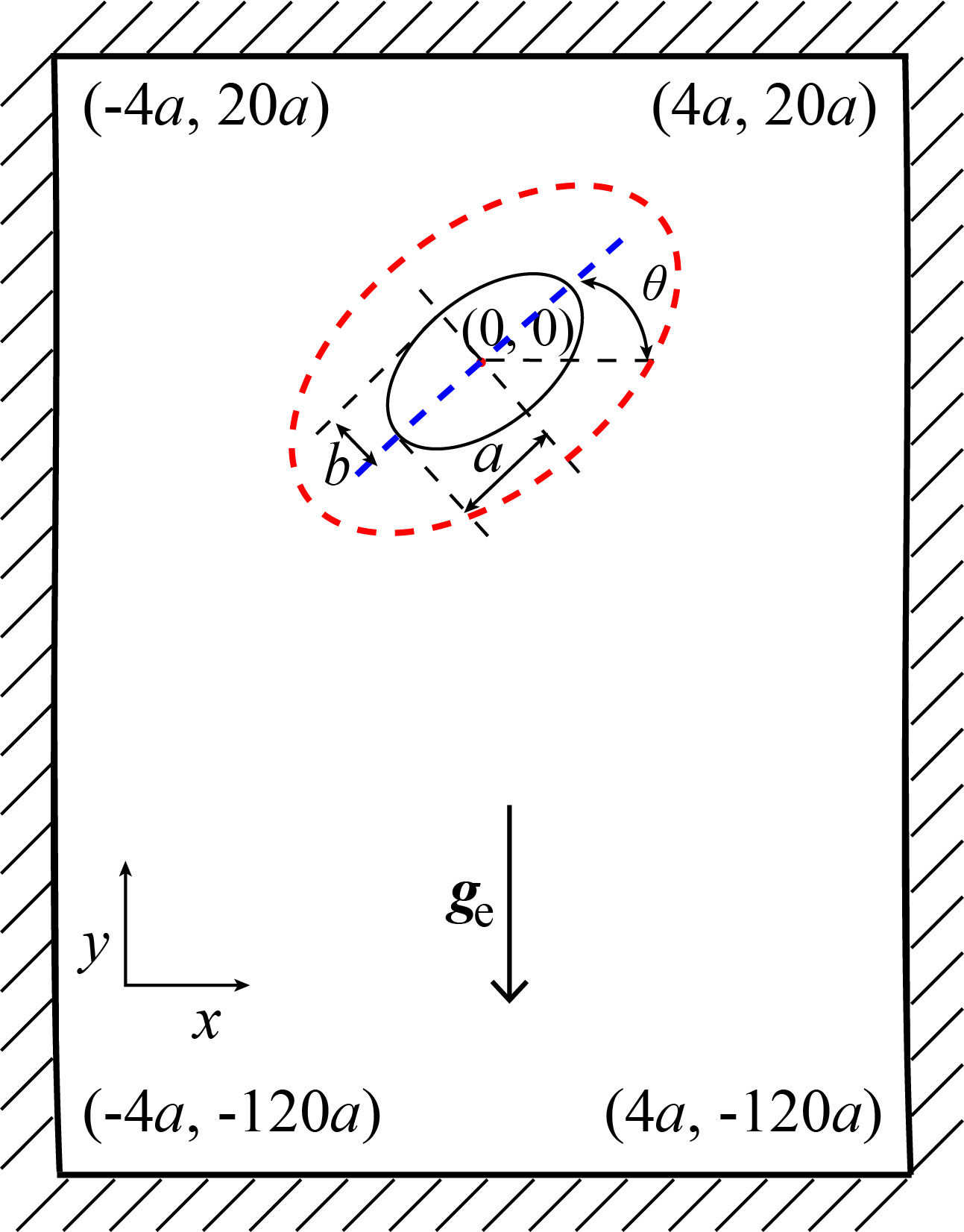}}\quad\quad\quad\quad
\subfloat[]{\includegraphics[width=6cm,valign=m]{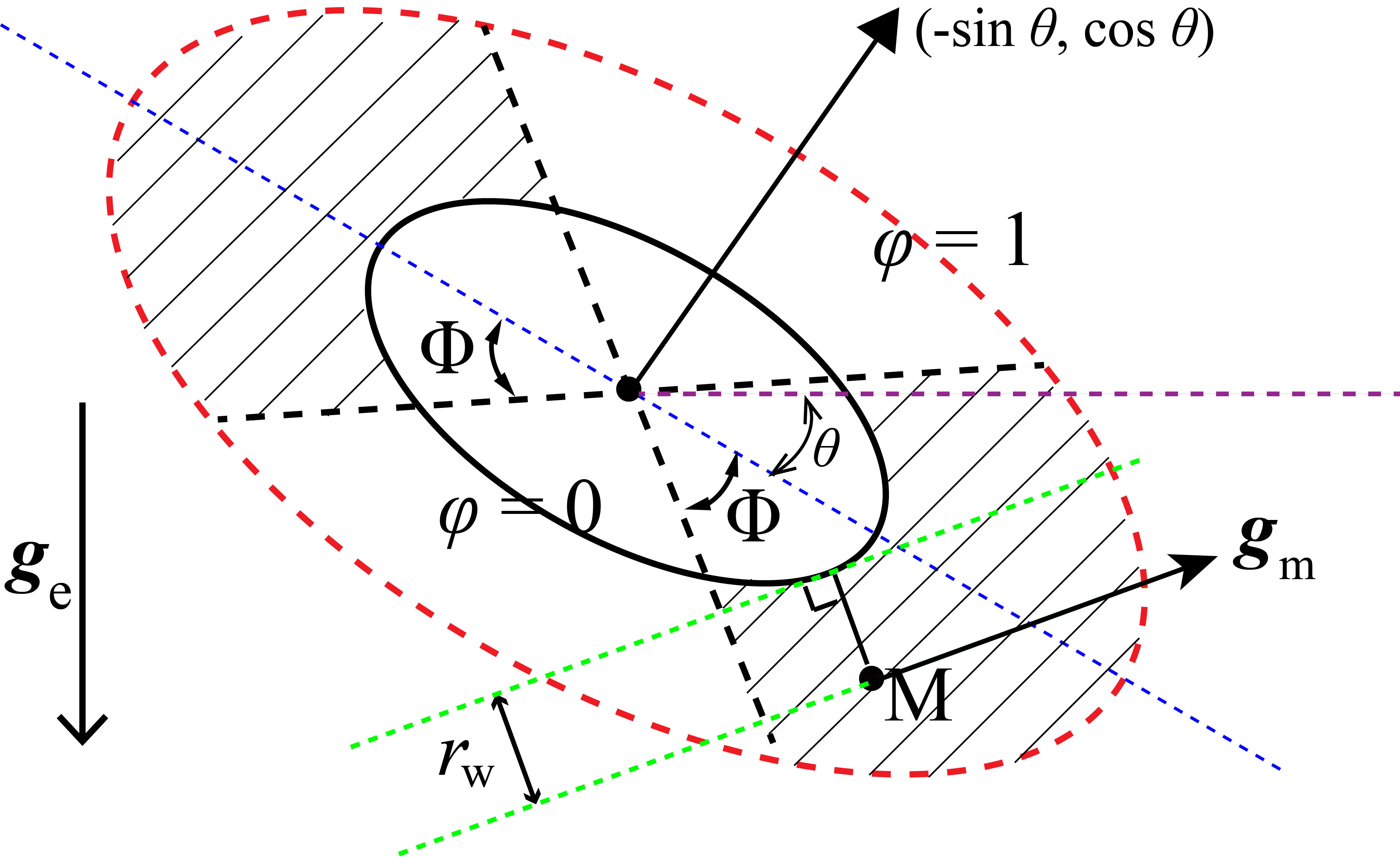}}\\
\caption{Schematic of electromagnetic control of a sedimenting elliptical particle. (a) Whole domain with the particle initially centered at $(0,0)$; (b) zoom in near the region of the particle during sedimentation.}
\label{fig:Lorenz_schematic}
\end{figure}

Fig.~\ref{fig:Lorenz_schematic} is the schematic of the electromagnetic force applied on the exterior fluid of an elliptical particle. The semi-major and semi-minor axes of the particle are defined as $a$ and $b$, respectively. The magnitude of the electromagnetic force on a fluid position M near the surface of the elliptical particle is 
\begin{equation}
g_{\text{m}}=e^{-\alpha_\text{m} r_\text{w}/2a}\beta_{\text{m}},
\end{equation}
in which $r_\text{w}$ is the distance between M and the corresponding projection point on the surface and is calculated via the signed distance function $\varphi (\bm{x})$~\cite{QIN2020109807}. $\alpha_\text{m}$ is a constant showing the electromagnetic penetration
into the flow which can be controlled by electrode spacing~\cite{chen2005active}. In our present study, $\alpha_\text{m}$ is chosen as 5. $\beta_{\text{m}}=N_{\text{m}}\frac{\nu^2\rho_f}{(2a)^3}$, in which $N_{\text{m}}$ is the dimensionless strength of the electromagnetic force.

The setup of the problem studied within this paper is shown in Fig.~\ref{fig:Lorenz_schematic}. The electromagnetic force $\bm{g}_{\text{m}}$ is parallel to the tangential direction of the interface. By using the signed distance function, the normal direction $\bm{n}=(n_x,n_y)$ of a fluid point M to the interface in two spatial dimensions can be obtained (Eq.~(\ref{eq:normal_in_jump})). Then the direction of the electromagnetic force is perpendicular to $\bm{n}$ and can be written as
\begin{equation}
\bm{\tau}=(-n_y,n_x).
\end{equation}
Finally, the density of the electromagnetic force can be calculated via
\begin{equation}
\bm{g}_{\text{m}}=\bm{\tau}\cdot g_{\text{m}}.
\end{equation}
It should be mentioned that the electromagnetic force is applied in a fluid region of
\begin{equation}
0\leq \varphi(\bm{x},t)\leq 1,
\end{equation}
in which $\varphi (\bm{x},t)$ is the level set function used in the immersed-interface method. Moreover, as shown in Fig.~\ref{fig:Lorenz_schematic}, the electromagnetic force only exists in a portion of the above fluid region with $\Phi=\frac{\pi}{4}$. 

It should be mentioned that the key idea of the electromagnetic force control is to suppress vortex generation and the adverse pressure gradient in order to reduce the periodic vortex shedding in the wake of the elliptical particle. Therefore, the direction of electromagnetic force should be in the opposite direction of the adverse pressure gradient. This requires
\begin{equation}
(-n_y,\,n_x)\cdot(-\sin\theta,\,\cos\theta)\geq 0,
\end{equation}
in which $\theta$ is instantaneous orientation angle of the elliptical particle. Here, $(-\sin\theta,\,\cos\theta)$ is the direction of the minor axis of the elliptical particle and the intersection angle between it and the gravity should not be less than $\frac{\pi}{2}$.

%\FloatBarrier

\subsection{Grid convergence study}
\label{sec:grid_con}
Electromagnetic control of the sedimentation of an elliptical particle is studied in this paper. Without the external electromagnetic force, the fluid-structure interaction (FSI) problem was proposed by Xia~et~al.~\cite{xia2009flow}. As shown in Fig.~\ref{fig:Lorenz_schematic}(a), the computational domain is chosen to be $[-4a,4a]\times[-120a,20a]$ and the elliptical particle initially centered at $(0,0)$ with an initial orientation angle of $\frac{\pi}{4}$. All boundaries of the outer domain are solid walls with Dirichlet boundary conditions. As a result of the gravity force, the elliptical \textcolor{black}{particle} sediments downwards. The viscous force and pressure on the particle surface drive it to move in the horizontal direction. Moreover, rotation of the particle is generated because of fluid torque on the particle. The density ratio between the elliptical particle and the fluid is represented by $\rho^*$ and is chosen as 1.5. The kinematic viscosity is $\tilde{\nu}=0.01\,\text{m}^2/\text{s}$, and the gravity acceleration is $\tilde{\bm{g}}_\text{e}=(0,-9.8\,\text{m}/\text{s}^2)$, in which the `tilde' symbol represents the physical units. The semi-major and semi-minor axes of the elliptical particle are set to $\tilde{a}=0.025\times10^{-2}\,\text{m}$ and $\tilde{b}=\frac{1}{2}\tilde{a}$, respectively. The II-LBM used in this study has been previously validated to be able to study this FSI problem without external electromagnetic forces~\cite{QIN2020109807}.

\begin{figure}[htb]
\centering
% \begin{subfigmatrix}{2} % number of columns
  \subfloat[]{\includegraphics[width=7.5cm]{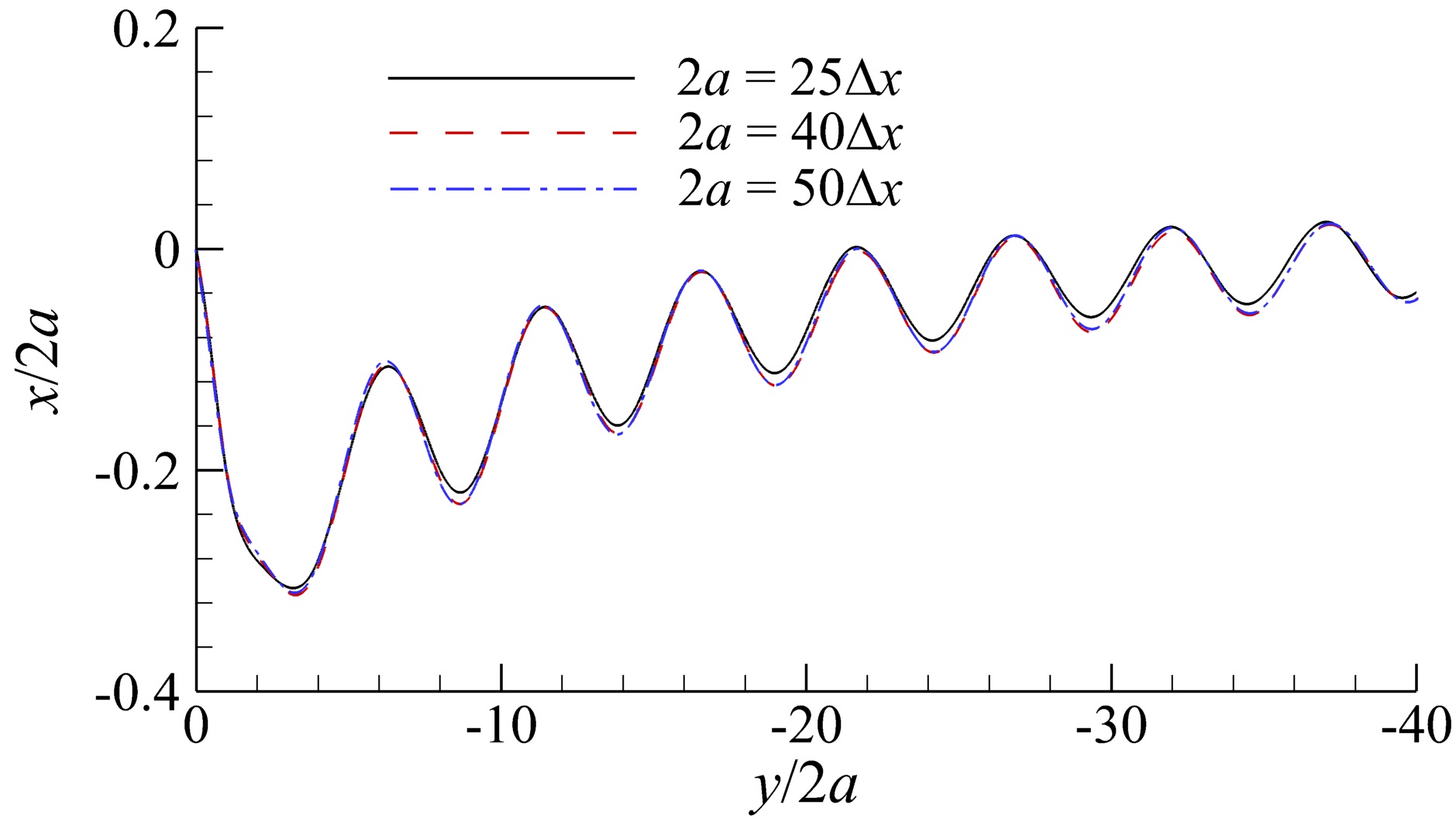}}
  \subfloat[]{\includegraphics[width=7.5cm]{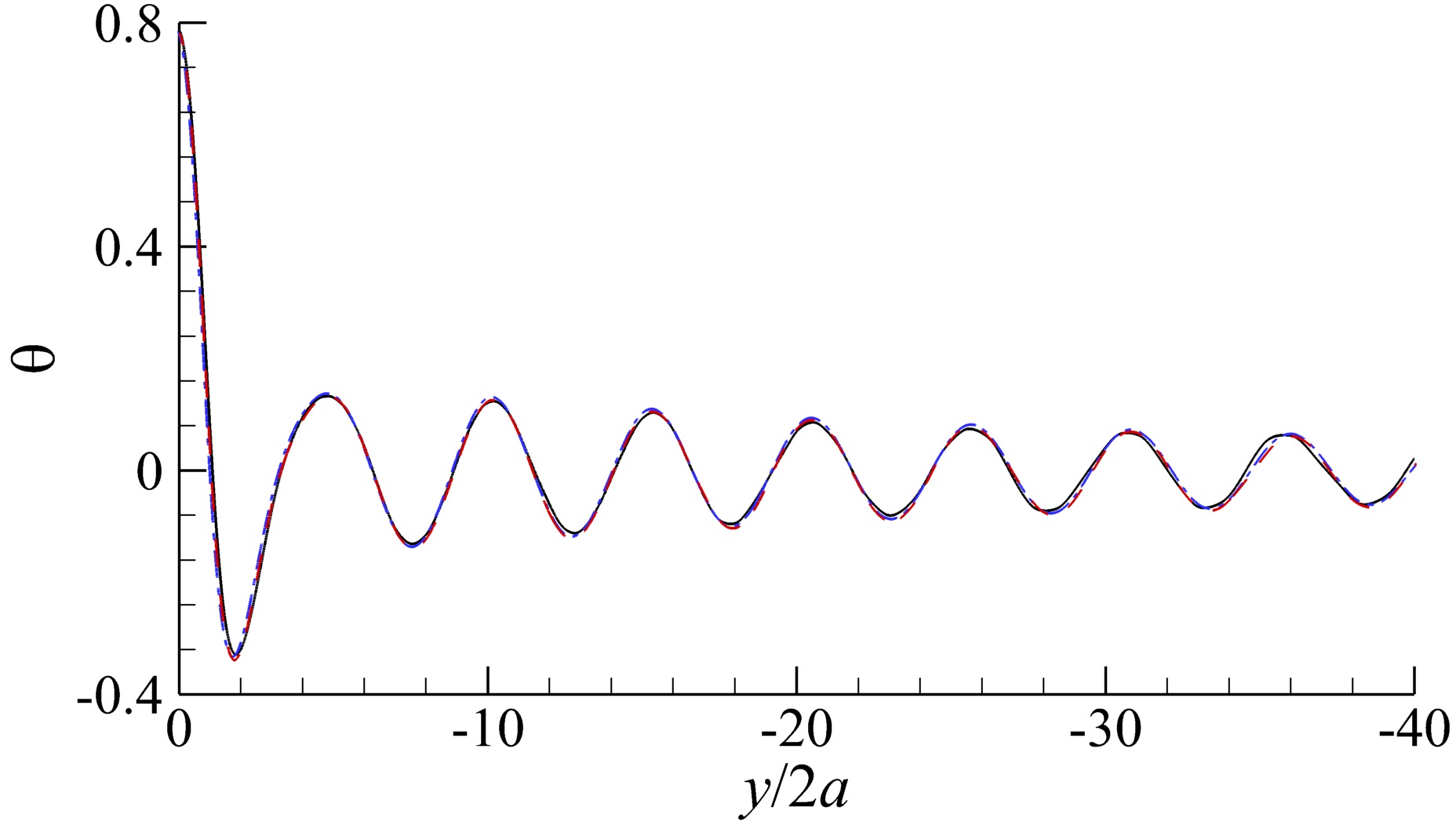}}
% \end{subfigmatrix}
 \caption{(a) Trajectories of the elliptical particle mass center, and (b) the orientation angle of the particle against the displacement of the particle mass center in the $y$ direction for three meshes with $N_\text{m}=6\text{e}3$ and $\rho^*=1.5$.}
\label{fig:iim_mag_Sedimentation_refine}
\end{figure}

To check the mesh size on the study of electromagnetic control of the particle sedimentation, simulation results for $N_\text{m}=6\text{e}3$ using three different meshes of $2a=25\Delta x$, $40\Delta x$ and $50\Delta x$ are compared. Fig.~\ref{fig:iim_mag_Sedimentation_refine} shows the trajectories of the elliptical particle and the relationship between the orientation angle and the displacement of the elliptical particle for different meshes. 
The two curves shown in Fig.~\ref{fig:iim_mag_Sedimentation_refine} are both periodic and are very close for three different meshes. However, the result obtained by $2a=25\Delta x$ shows larger differences between the latter two meshes. Considering the accuracy of result and the computational cost, we choose $2a=40\Delta x$ to study the electromagnetic control of the particle sedimentation.

\section{Results and discussions}
\subsection{\textcolor{black}{Sedimentation of an elliptical particle}}
\begin{figure}[htb!]
\centering
% \begin{subfigmatrix}{2} % number of columns
\includegraphics[width=15cm]{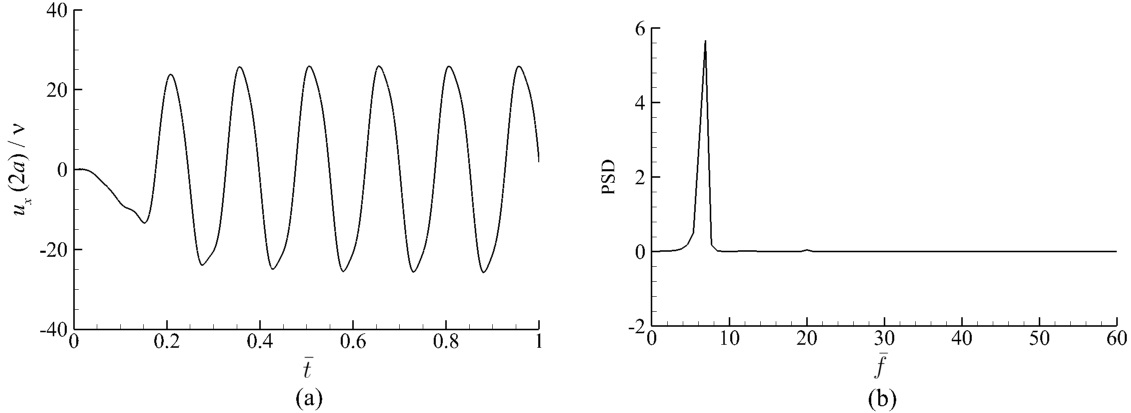}
% \end{subfigmatrix}
\caption{\textcolor{black}{(a) $u_x$ of the characteristic point C against the dimensionless time, and (b) power spectral density of the frequency for $u_x$ of point C.}}
\label{fig:characteristic_freq}
\end{figure}

\begin{figure}[htb!]
\centering
\includegraphics[width=16cm,valign=t]{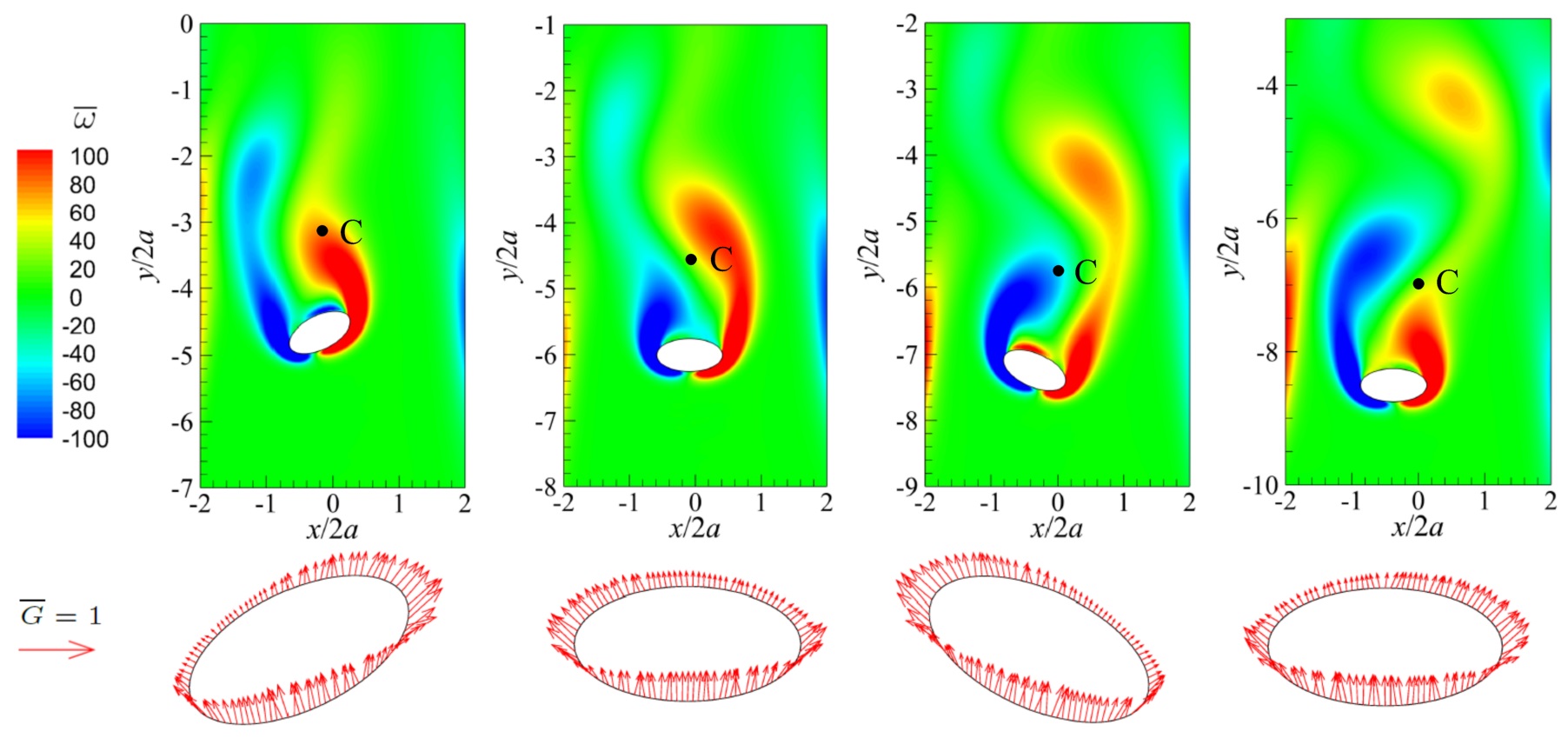}
\caption{Snapshots of vorticity fields (upper) and Lagrangian forces on the particle boundary (lower) for a sedimenting ellpitical particle. From left to right, the four typical snapshots correspond to $\bar{t}_1$, $\bar{t}_2$, $\bar{t}_3$, and $\bar{t}_4$ in Fig.~\ref{fig:dis_ang_1.5}(b), respectively.}
\label{fig:NO_mag_force_vor}
\end{figure}

\textcolor{black}{Without the electromagnetic force, the simple sedimentation problem will result in an oscillating motion of the particle~\cite{xia2009flow}. The problem setup is the same as in Sec.~\ref{sec:grid_con} with $\rho^*=1.5$ except that $N_\text{m}$ is set to zero here. As shown later in Figs.~\ref{fig:dis_ang_1.5}\subref{fig:40dx_dis_mag_5} and \ref{fig:dis_ang_1.5}\subref{fig:40dx_ang_mag_5}, both the trajectory and the time response of the orientation angle of the particle are periodic. The density ratio is within a relatively high range $((\rho^*-1)\sim O(1))$ where the convection of the fluid plays an important role in the terminal velocity ($U_\text{t}$) of the particle~\cite{xia2009flow}. To define the feature of the flow, the Reynolds number and the characteristic frequency of the flow should be specified. The Reynolds number is defined via 
\begin{equation}
Re=\frac{U_\text{t}\,2a}{\nu}.
\label{eq:re}
\end{equation}
Within the high density ratio range, the Reynolds number scales as $Re\sim (\rho^*-1)^{0.5}$. In our present study, the resulting $Re$ calculated by the II-LBM for $\rho^*=1.5$ is 32.7. The characteristic frequency of the particle during sedimentation can be obtained by performing a fast Fourier transform of $u_x$ for a fixed position relative to the particle mass center in the $y$ direction in the wake, in which $u_x$ is the fluid velocity in the $x$ direction. Here, the position is chosen as $(0,\,X_{\text{c},y}(t)+3a)$ (point C in Fig.~\ref{fig:NO_mag_force_vor}). $u_x$ for point C against the dimensionless time $\bar{t}$ is presented in Fig.~\ref{fig:characteristic_freq}(a), in which $\bar{t}$ is the dimensionless time and is defined as
\begin{equation}
\overline{t}=\frac{t}{(2a)^2/\nu}.
\end{equation}
The dimensionless characteristic frequency can be defined as $\bar{f}=\frac{f\,(2a)^2}{\nu}$, where $f$ is the frequency of vortex shedding. As shown in Fig.~\ref{fig:characteristic_freq}(b), the peak of the power spectral density (PSD) corresponds to $\bar{f}=$6.93.}

Four snapshots of vorticity fields and force distributions on the surface of the elliptical particle are shown in Fig.~\ref{fig:NO_mag_force_vor}, in which the dimensionless vorticity is defined as
\begin{equation}
\overline{\omega} = \frac{\omega}{\nu/(2a)^2},
\end{equation}
where $\omega$ is the vorticity, and the dimensionless Lagrangian force is written as $\overline{\bm{G}}=\frac{\bm{G}}{10^5\,\nu^2\rho_\text{f}/(2a)^3}$. One can see that vortex shedding causes uneven vorticities attached to the surface of the particle and uneven forces on the left and right sides of the surface. The shedding \textcolor{black}{vortices} are generated from flow separation when the boundary layer has travelled in an adverse pressure gradient for a long distance to result in a relative zero velocity on the surface. Because of the forces and resulting torques, the particle will rotate during the sedimentation process.

\subsection{Mechanism of the electromagnetic flow control}
\begin{figure}[htb!]
\centering
\includegraphics[width=6.5cm]{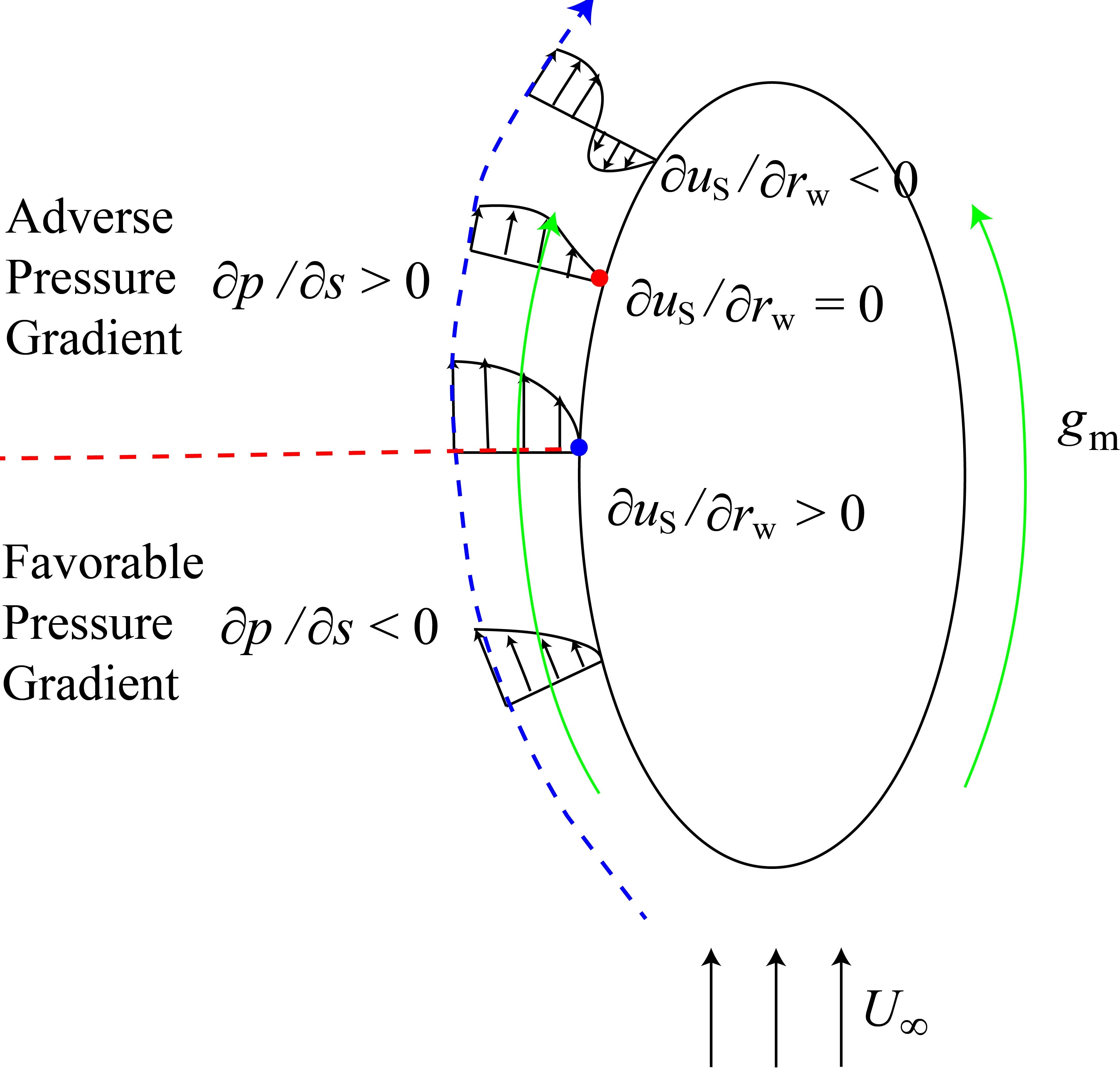}
\caption{Schematic of the boundary layer controlled by the electromagnetic force. The blue and red points represent the zero pressure gradient point and the stagnation point, respectively.}
\label{fig:boundary_layer}
\end{figure}

Electromagnetic control of a sedimenting elliptical particle is studied in this section. The rotation of the particle during sedimentation is caused by vortex shedding. Let us explain why the electromagnetic force introduced in Sec.~\ref{sec:magnetic} can control vortex shedding so that the rotational motion of the particle can be controlled. The mechanism behind the electromagnetic control of vortex shedding is that the vorticity generation and the adverse pressure gradient on the boundary layer of the particle can both be suppressed by this kind of electromagnetic forces. Fig.~\ref{fig:boundary_layer} shows the schematic of the boundary layer for uniform flow around an elliptical cylinder.

The vorticity equation can be written as
\begin{equation}
\begin{aligned}
\frac{\text{D}\bm{\omega}}{\text{D}t}
%&=\frac{\partial\omega}{\partial t}+(\bm{u}\cdot \nabla)\omega \\&
=(\bm{\omega}\cdot\nabla)\bm{u}-\bm{\omega}(\nabla\cdot\bm{u})+\frac{1}{\rho_\text{f}^2}\nabla\rho_\text{f}\times\nabla p+
\nabla\times\frac{\nabla\cdot\bm{\tau}}{\rho_\text{f}}+\nabla\times\frac{\bm{g}_\text{m}}{\rho_\text{f}},
\end{aligned}
\label{eq:vorticity_eq}
\end{equation}
in which the first term on the right hand side represents vortex stretching and should be zero in two spatial dimensions. For an incompressible fluid, the second and third terms on the right hand side of Eq.~(\ref{eq:vorticity_eq}) are also zero. The viscous force on the boundary layer which is generated by the motion of the particle obeys the Newton's viscous law and is inversely proportional to $\frac{\text{d}\bm{u}}{\text{d}r_\text{w}}$. Therefore, the direction of the viscous force is in contrast to the electromagnetic force. For this reason, the vorticity generation on the boundary layer in front of the separation point (the red point in Fig.~\ref{fig:boundary_layer}) can be suppressed by the electromagnetic force. It should be noted that when the electromagnetic force is stronger than the viscous force, the magnetic induced vorticity becomes the dominant vorticity in the fluid. It can be concluded that appropriate electromagnetic forces can suppress vorticity generation to attenuate the energy so that the rotational motion of the particle can be suppressed.

Because vortex shedding is caused by the adverse pressure gradient, a suppression of the adverse pressure gradient leads to a suppression of vortex shedding. Within the boundary layer, the streamwise momentum equation can be approximately stated as 
\begin{equation}
u_\text{S}\frac{\partial u_\text{S}}{\partial s}=-\frac{1}{\rho_\text{f}}\frac{\text{d}p}{\text{d}s}+\nu\frac{\partial^2u_\text{S}}{\partial n_\text{c}^2}+g_\text{m},
\end{equation}
in which $u_\text{S}$ is the streamwise velocity, and $s$ and $n_\text{c}$ denote streamwise and normal coordinates. Without the effect of the electromagnetic force, a strong enough adverse pressure gradient (i.e. $\frac{\text{d}p}{\text{d}s}>0$) will cause $u_\text{S}$ to decrease along $s$ and goes to zero, resulting in the occurrence of flow separation. Due to the fact that the directions of the electromagnetic force and the adverse pressure gradient are opposite to each other, the electromagnetic force will suppress the adverse pressure gradient.

\subsection{Effect of the electromagnetic force strength}
\begin{figure}[htb]
\centering
\includegraphics[width=0.9cm,valign=t]{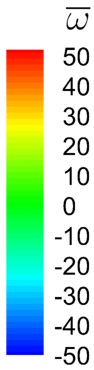}
\includegraphics[width=8cm,valign=t]{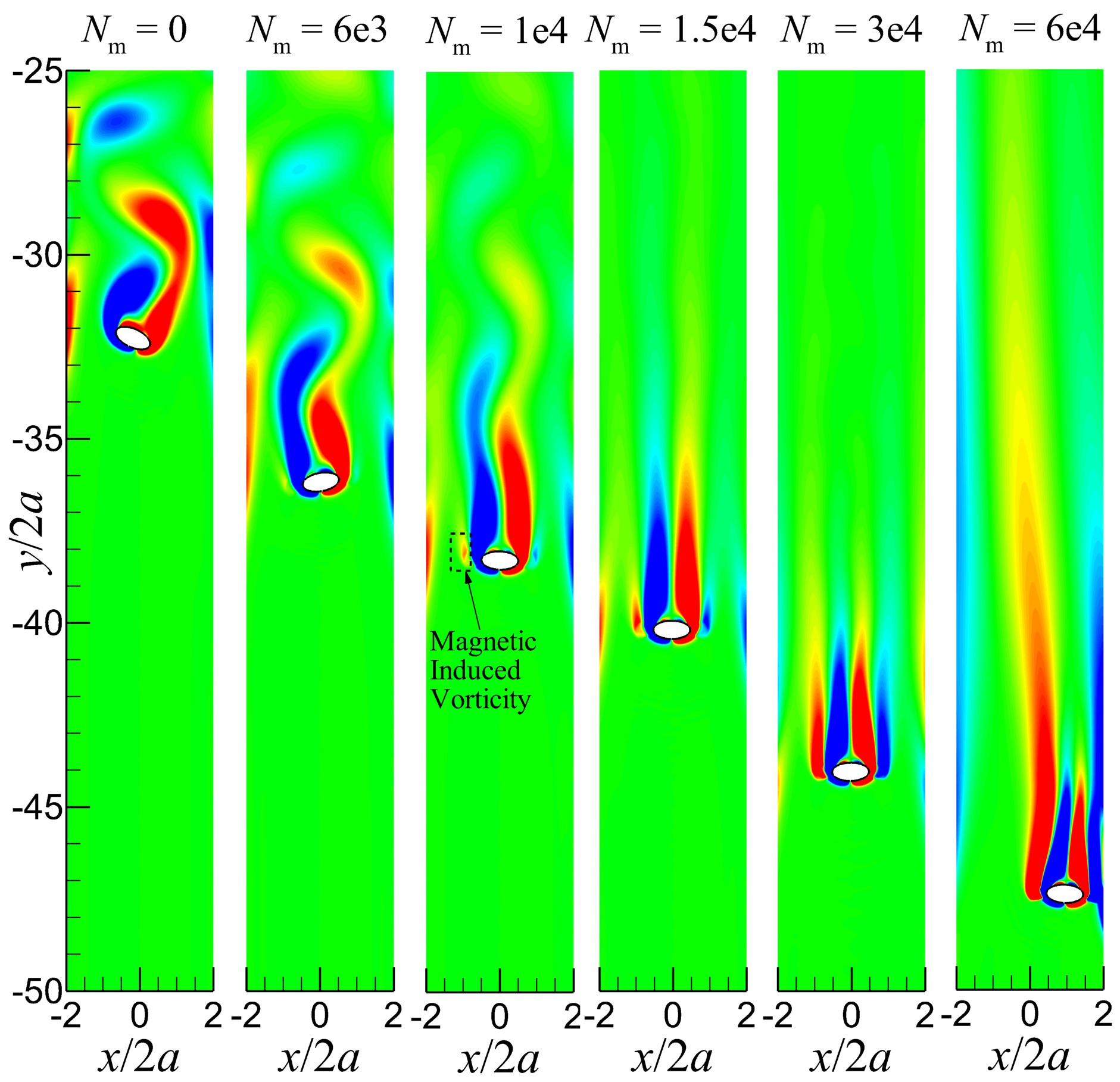}
\caption{Vorticity fields for different magnitudes of electromagnetic forces at $\bar{t}=1$.}
\label{fig:Mag_vor_whole}
\end{figure}

Fig.~\ref{fig:Mag_vor_whole} shows the vorticity field of a sedimenting elliptical particle under different magnitudes of electromagnetic forces at $\bar{t}=1$. As the magnitude of the electromagnetic force increases, the instantaneous displacement of the particle mass center in the $y$ direction ($\frac{y}{2a}$) are -32.04, -35.6, -37.44, -39.12, -42.28 and -43.60, respectively. This indicates that the external electromagnetic force accelerates the sedimentation. In the range of $0\leq N_\text{m} \leq 1.5\text{e}4$, as the electromagnetic force increases, the vorticity on the boundary layer becomes symmetric and vortex shedding vanishes. It should be mentioned that, as the electromagnetic force grows to $N_\text{m} = 1\text{e}4$, the vorticity induced by the electromagnetic force appears. When the electromagnetic force continuously increases, the intensity of the induced vorticity also increases. In the case of $N_\text{m} = 3\text{e}4$, the intensity of the induced vorticity is comparable to the intensity of the vorticity on the boundary layer. In the case of $N_\text{m} = 6\text{e}4$, the induced vorticities are stronger than the vorticities on the boundary layer. Therefore, in the range of $0\leq N_\text{m} \leq 3\text{e}4$, larger electromagnetic forces stabilize wakes behind the particle. However, when the electromagnetic is excessive (e.g. in the case of $N_\text{m} = 6\text{e}4$), the particle's center of mass deviate from the center of side walls because of this electromagnetic force. Therefore, only electromagnetic forces of appropriate strength can stabilize the motion of the particle. 

%\FloatBarrier

\begin{figure}[htb]
\centering
\includegraphics[width=16cm]{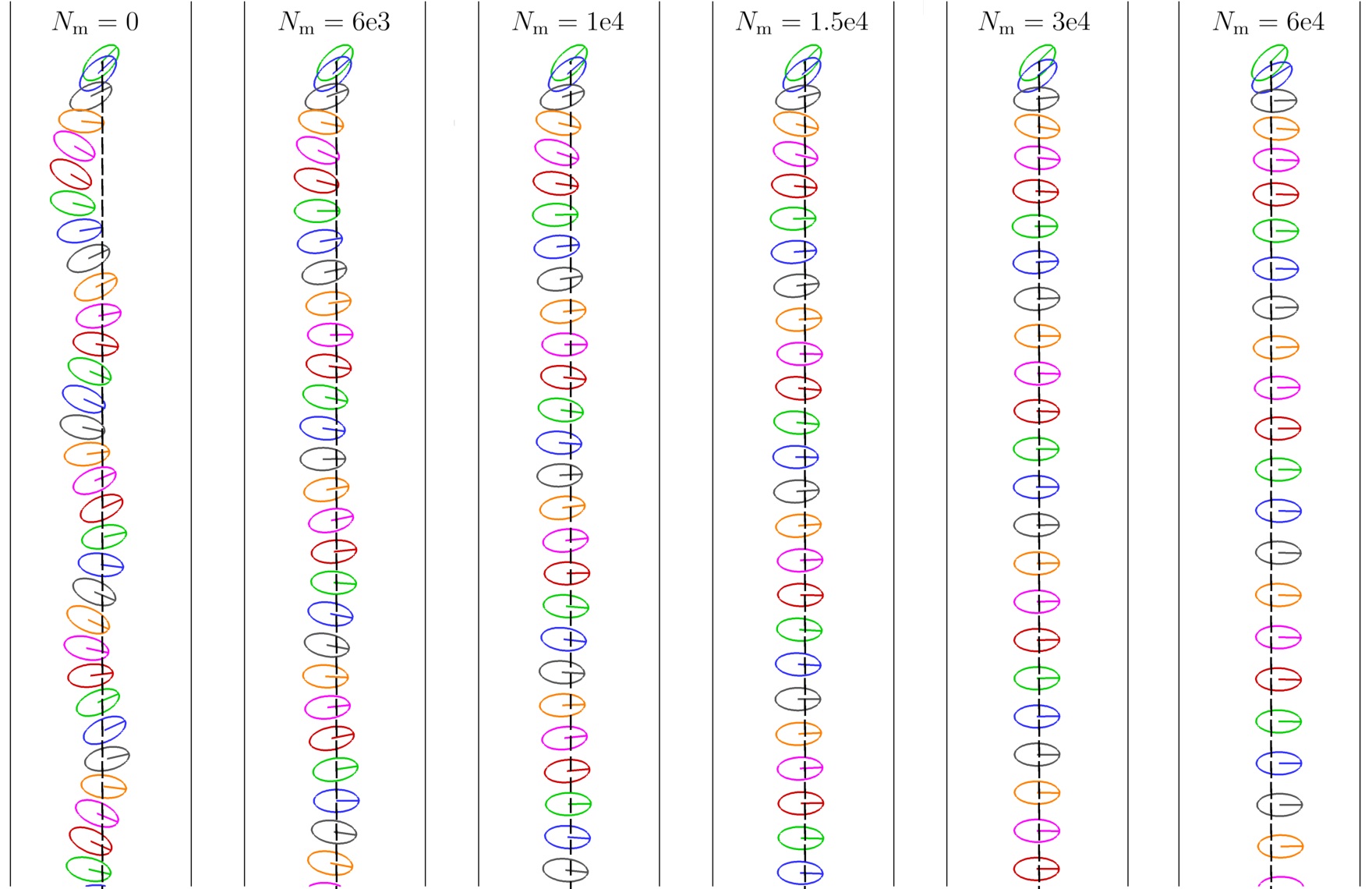}
\caption{\textcolor{black}{Sedimentation processes for the elliptical particles under the the effect of different magnitudes of electromagnetic forces for $\rho^*=1.5$.}}
\label{fig:angle_dyn}
\end{figure}

\begin{figure}[htb]
\centering
\subfloat[]{\label{fig:40dx_dis_mag_5}\includegraphics[height=4.5cm]{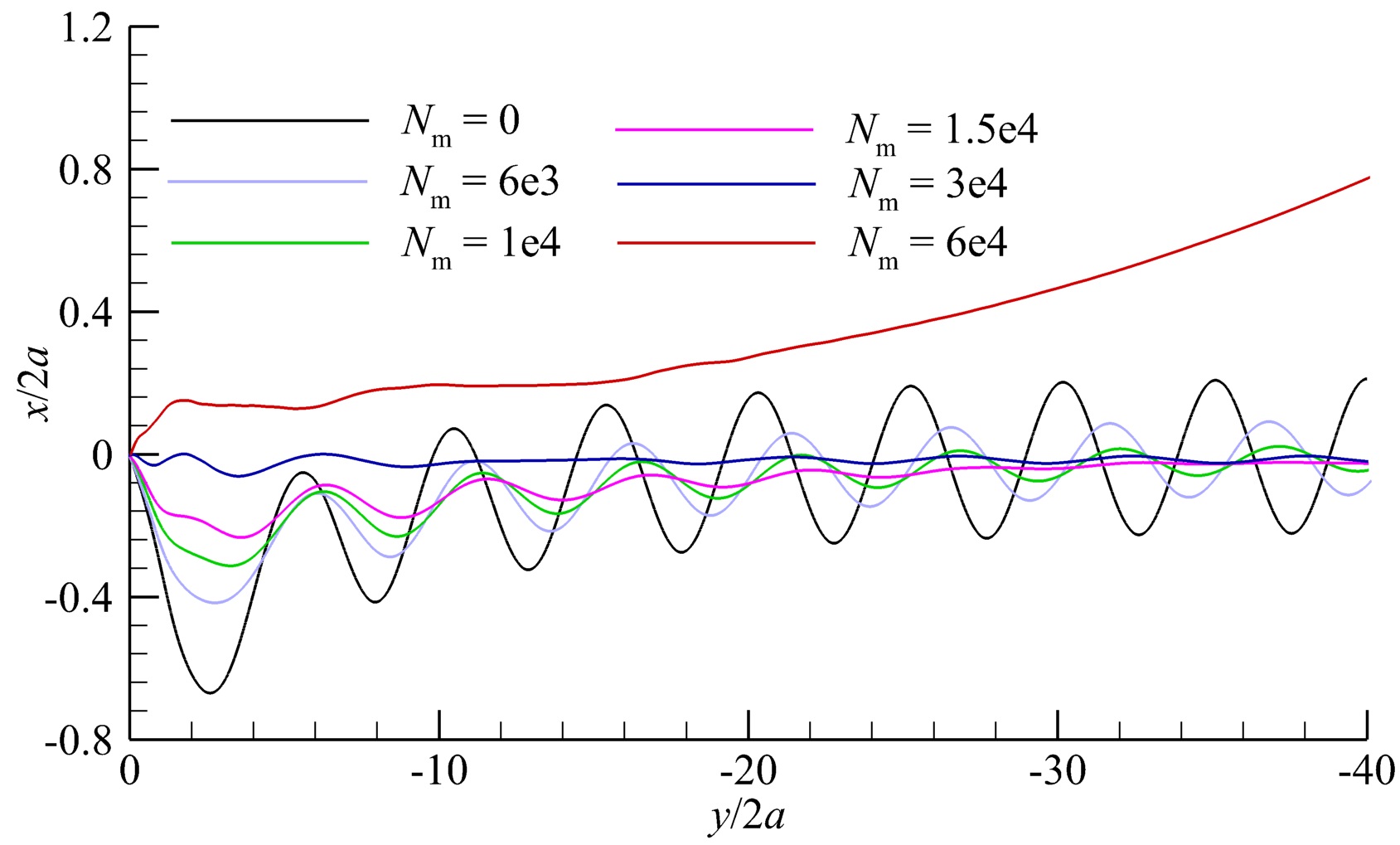}}\,\,
\subfloat[]{\label{fig:40dx_ang_mag_5}\includegraphics[height=4.5cm]{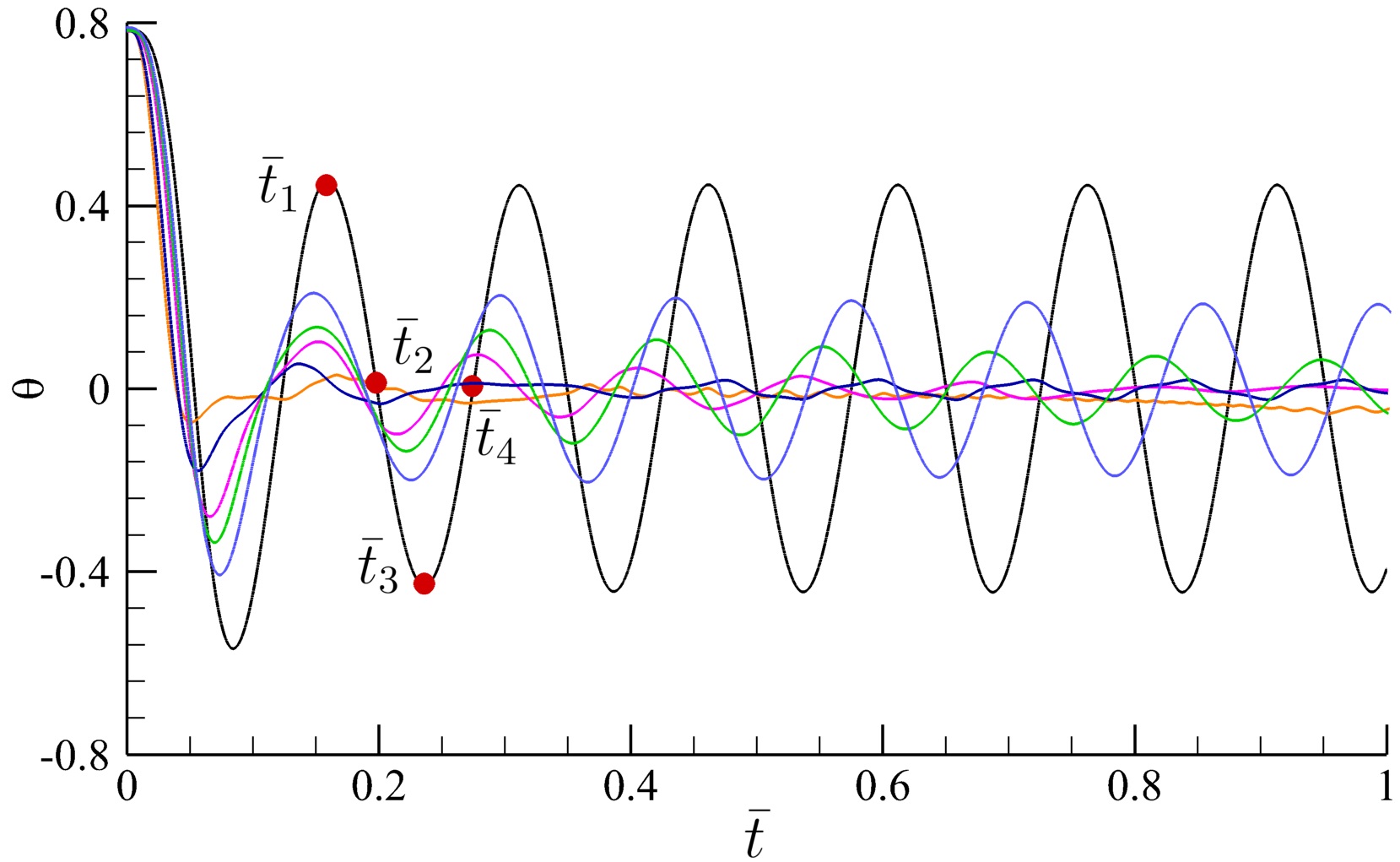}}
\caption{(a) Trajectories of the particle mass center during sedimentation, and (b) Orientation angle \textcolor{black}{against the dimensionless time} for different magnitudes of electromagnetic forces.}
\label{fig:dis_ang_1.5}
\end{figure}

\textcolor{black}{To further demonstrate the effect of electromagnetic forces on the sedimenting process, Fig.~\ref{fig:angle_dyn} shows the motion of the particles for different $N_\text{m}$. A detailed comparison of the trajectories of the sedimentation for different $N_\text{m}$ is shown in Fig.~\ref{fig:dis_ang_1.5}\subref{fig:40dx_dis_mag_5}.} Without the electromagnetic force (i.e. $N_\text{m}=0$), the elliptical particle flutters downwards and the particle mass center is around the middle of the side walls~\cite{xia2009flow}. This is consistent with the results reported by Huang et al.~\cite{huang_hu_joseph_1998} that a settling ellipse will turn its major axis to the vertical direction for sufficiently small Reynolds numbers. For $N_\text{m}=6\text{e}3$, the particle still flutters, although the oscillation amplitude in the $x$ direction becomes smaller. As the electromagnetic force increases, the oscillation amplitude in the $x$ direction becomes less and less until $N_\text{m}=3\text{e}4$. For $N_\text{m}=3\text{e}4$, in a initial range of $-2\leq y/8a\leq 0$, the elliptical particle oscillates with a small amplitude in the $x$ direction. Subsequently, the displacement in the $x$ direction is almost zero.

\begin{figure}[htb]
\centering
\includegraphics[width=15cm]{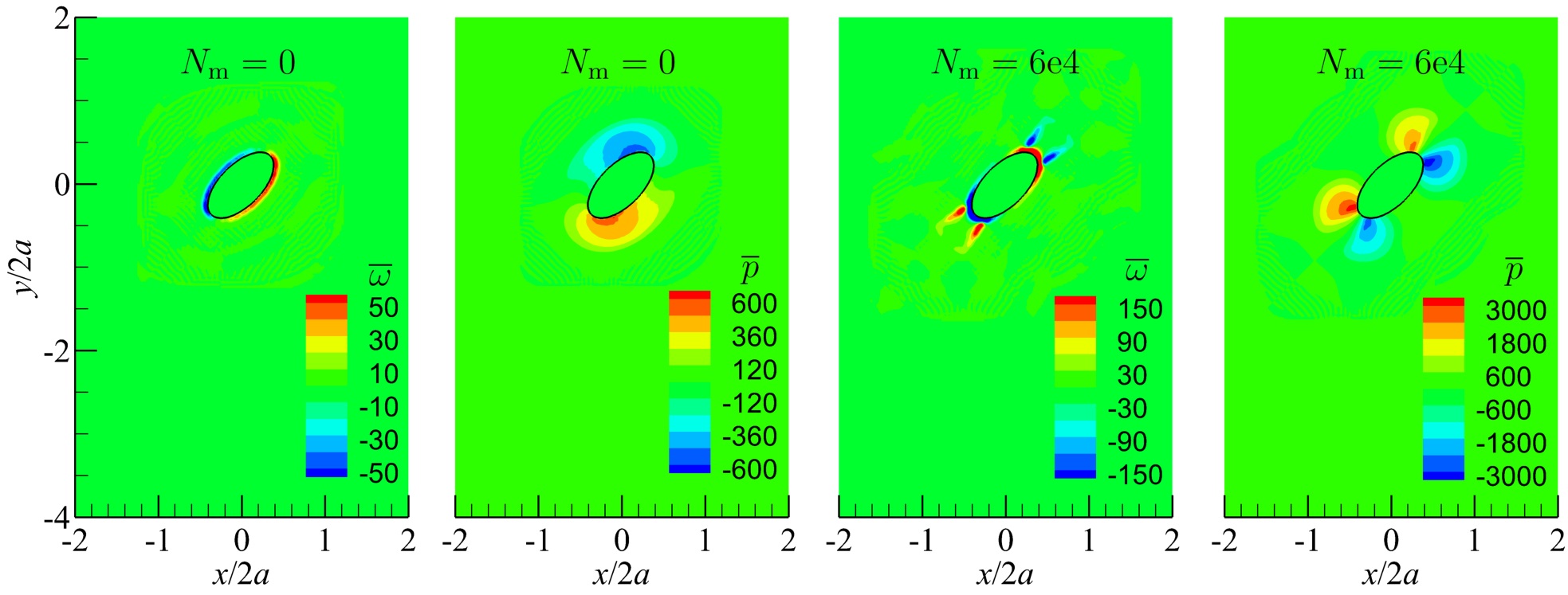}
\caption{Vorticity and pressure fields for $N_\text{m}=0$ and $N_\text{m}=6\text{e}4$ at $\bar{t}=0.002$.}
\label{fig:vor_pre}
\end{figure}

Interestingly, in the case of $N_\text{m}=6\text{e}4$, the elliptical particle moves towards the right wall of the domain. Feng~et~al. \cite{feng1994direct} reported that a difference between sedimentations of a circular particle and an elliptical particle at a high enough Reynolds number is that the circular particle may achieve its lateral equilibrium position off the center of the side walls, stabilized by a unidirectional rotation. However, a stabilizing pressure couple does not allow the ellipse to rotate similarly, resulting in the ellipse to oscillate about the center of the side walls. Because the viscous force on the boundary layer is well balanced by the electromagnetic force for the $N_\text{m}=6\text{e}4$ case, the motion of the particle will not be delayed by the fluid. Therefore, both the rotational speed caused by the initial orientation angle and the translational speed caused by the gravity force are higher than other cases. This means the boundary layer forms at the initial stage of the sedimentation, and the lift force is insignificant compared to the pressure differences on the left and right side of the particle. As displayed in Fig.~\ref{fig:vor_pre}, the pressure differences tend to push the particle to the right wall, in which the dimensionless pressure is defined as
\begin{equation}
\overline{p}=\frac{p-p_0}{\nu^2/D^2}.
\end{equation}
Here $p_0$ is the initial pressure of the fluid. When the rotation of the particle caused by the initial orientation angle ends for $N_\text{m}=6\text{e}4$, the magnetic induced vorticities~\cite{Himo2018} will be uneven because the particle is deviated from the center of the side walls. The difference of the magnetic induced vorticities makes the particle move towards the right wall continuously.

For different magnitudes of electromagnetic forces, the orientation angle of the elliptical particle against time is shown in Fig.~\ref{fig:dis_ang_1.5}\subref{fig:40dx_ang_mag_5}. In the range of $0\leq N_\text{m} \leq 3\text{e}4$, the curve of $\theta$ against time is periodic. As the electromagnetic force increases, the oscillation amplitude of the orientation angle decreases. The specific case of $N_\text{m}=0$ is different from other cases in that the oscillation amplitude does not decrease against time. When $N_\text{m}>0$, the oscillation magnitude of $\theta$ decreases with the increase of the sedimenting distance. For $N_\text{m}\geq 3\text{e}4$, the oscillation amplitude tends to zero after the elliptical particle sediments a short distance. This means that the rotations of the elliptical particle are very weak during sedimentation because of the electromagnetic force.
%\begin{figure}[htb]
%\centering
%\includegraphics[width=9cm]{40dx_ang_mag_5}
%\caption{}
%\label{fig:40dx_ang_mag_5}
%\end{figure}
%\FloatBarrier

%\FloatBarrier

The total kinetic energy of the elliptical particle ($\text{KE}_\text{T}$) is defined as
\begin{equation}
\text{KE}_\text{T}=\text{KE}_x+\text{KE}_y+\text{KE}_\text{R},
\end{equation}
in which $\text{KE}_x$ and $\text{KE}_y$ are kinetic energies in the $x$ and $y$ directions, respectively. The kinetic energies are calculated via
\begin{equation}
\text{KE}_x=0.5\rho^*\frac{V_\text{s}U_{\text{c},x}^2}{2a\nu^2},
\end{equation}
\begin{equation}
\text{KE}_y=0.5\rho^*\frac{V_\text{s}U_{\text{c},y}^2}{2a\nu^2},
\end{equation}
in which $U_{\text{c},x}$ and $U_{\text{c},y}$ are velocities of the particle mass center in the $x$ and $y$ directions, respectively. $\text{KE}_\text{R}$ is the rotational energy defined by
\begin{equation}
\text{KE}_\text{R}=0.5\frac{\mathbb{I}_{zz}W_z^2}{2a\nu^2}.
\end{equation}

\begin{figure}[htb]
\centering
\subfloat[]{\label{fig:Energy_comversion}\includegraphics[width=7.5cm]{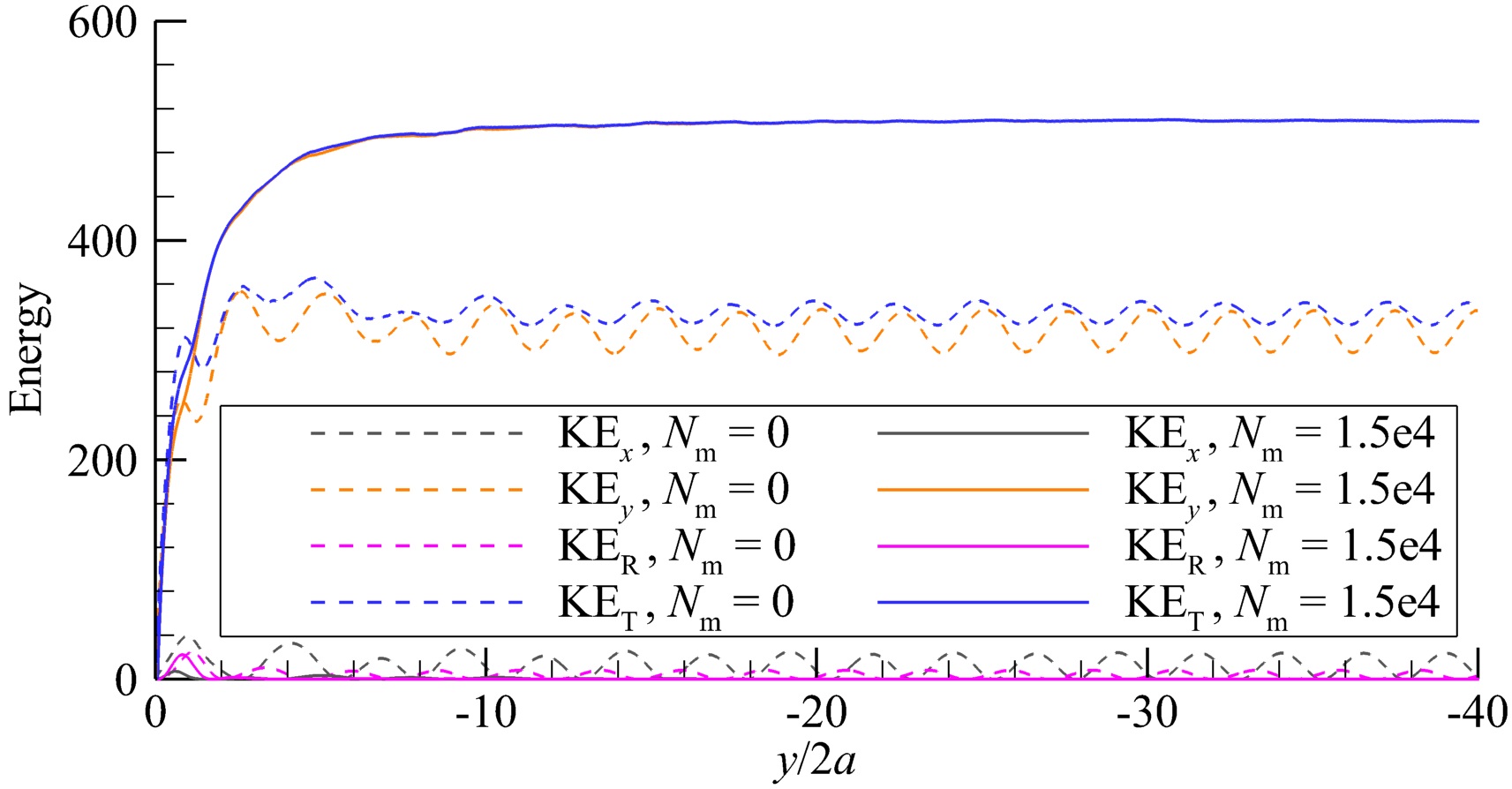}}\,\,
\subfloat[]{\label{fig:vel_all}\includegraphics[width=7.5cm]{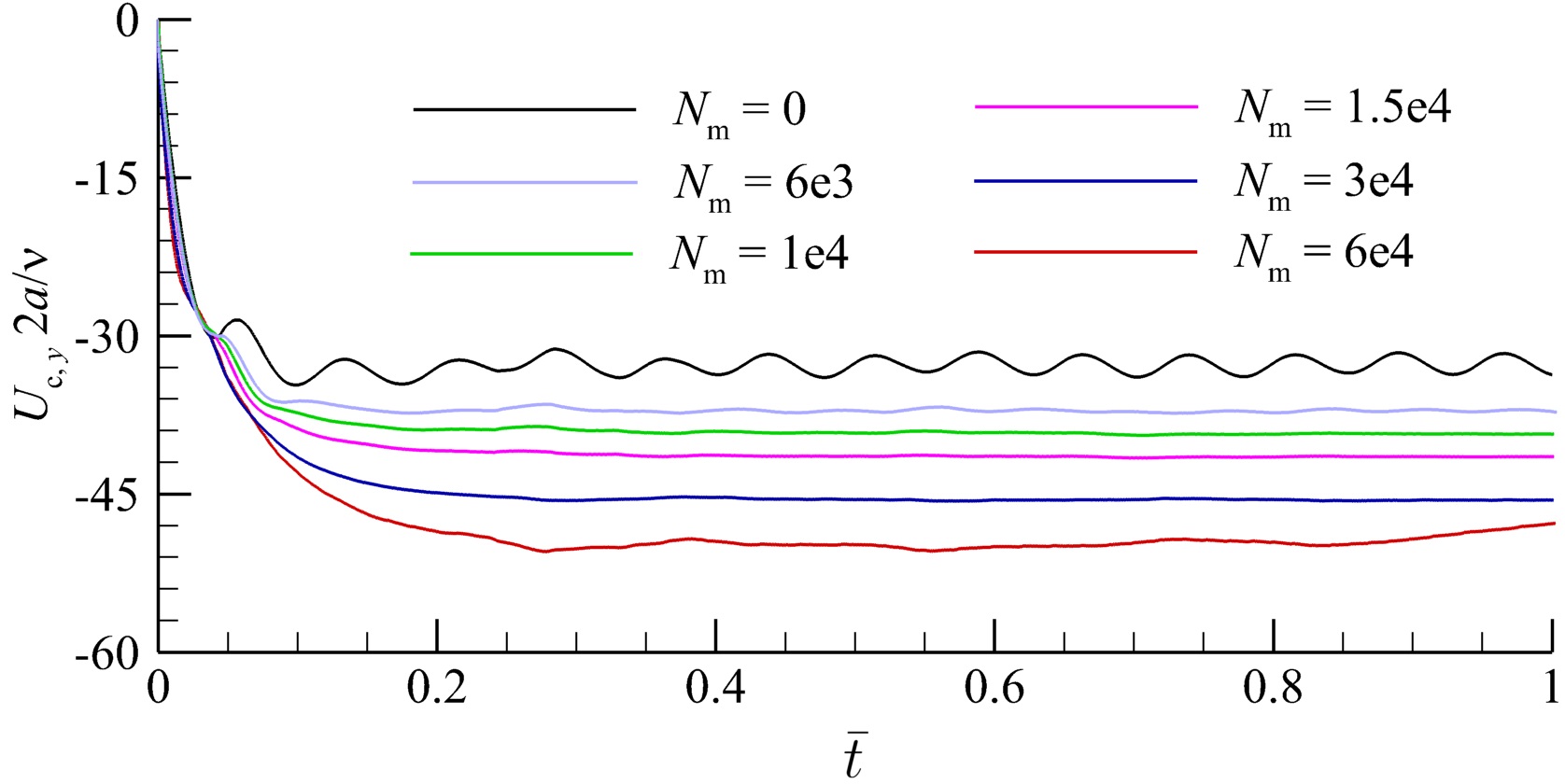}}
\caption{(a) Comparison of particle energies for $N_\text{m}=0$ and $N_\text{m}=1.5\text{e}4$ when $\rho^*=1.5$, and (b) \textcolor{black}{Velocity of the particle in the $y$ direction against the dimensionless time for different magnitudes of electromagnetic forces}.}
\label{fig:energy_vel_1.5}
\end{figure}

When sedimenting, the gravitational potential energy of the elliptical particle becomes smaller as the mass center of the particle is lower. Therefore, the $y$ displacement of particle center can be seen as equivalent to the potential energy of the particle. The gravitational potential energy is converted into the kinetic energy of the particle and the kinetic energy of the fluid. Fig.~\ref{fig:energy_vel_1.5}\subref{fig:Energy_comversion} shows the relationship between the $y$ displacement of particle center and the energies of the particle. For $-2\leq y/2a\leq 0$, the total energy of the particle is larger without the electromagnetic force than $N_\text{m}=1.5\text{e}4$. After this initial stage, the total energy of the particle becomes larger for the case of $N_\text{m}=1.5\text{e}4$, and the difference between $N_\text{m}=1.5\text{e}4$ and $N_\text{m}=0$ increases with the sediment distance. Because of the electromagnetic force, the kinetic energy in the $x$ direction and rotation energy become negligible, and the total energy of the particle is almost the same as the kinetic energy in the $y$ direction. The increase of the total energy and kinetic energy in the $y$ direction with the electromagnetic force indicates that the electromagnetic force is useful to reduce the energy loss during sedimentation. Therefore, the sedimentation distance increases when $N_\text{m}$ increases (see Fig.~\ref{fig:Mag_vor_whole}). When building a sedimenting device, it is also possible to accelerate the sedimentation of the device by adding the electromagnetic force. \textcolor{black}{From the evolution of the sedimentation velocity shown in Fig.~\ref{fig:energy_vel_1.5}\subref{fig:vel_all}, the periodic oscillation has been suppressed with the electromagnetic force. Moreover, the Reynolds number defined by Eq.~(\ref{eq:re}) is increased from 32.7 for $N_\text{m}=0$ to 45.4 for $N_\text{m}=3\text{e}4$. This means that the terminal velocity has gained 38.8\%.}

%\begin{figure}[htb]
%\centering
%\includegraphics[width=10cm]{40dx_vel_mag_5}
%\caption{\textcolor{black}{Velocity of the particle in the $y$ direction against the dimensionless time for different magnitudes of electromagnetic forces}.}
%\label{fig:vel_all}
%\end{figure}
%\FloatBarrier
\subsection{Effect of the initial orientation angle}
\begin{figure}[htb]
\centering
\includegraphics[width=1.cm,valign=t]{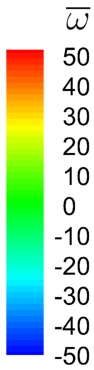}
\includegraphics[width=8.5cm,valign=t]{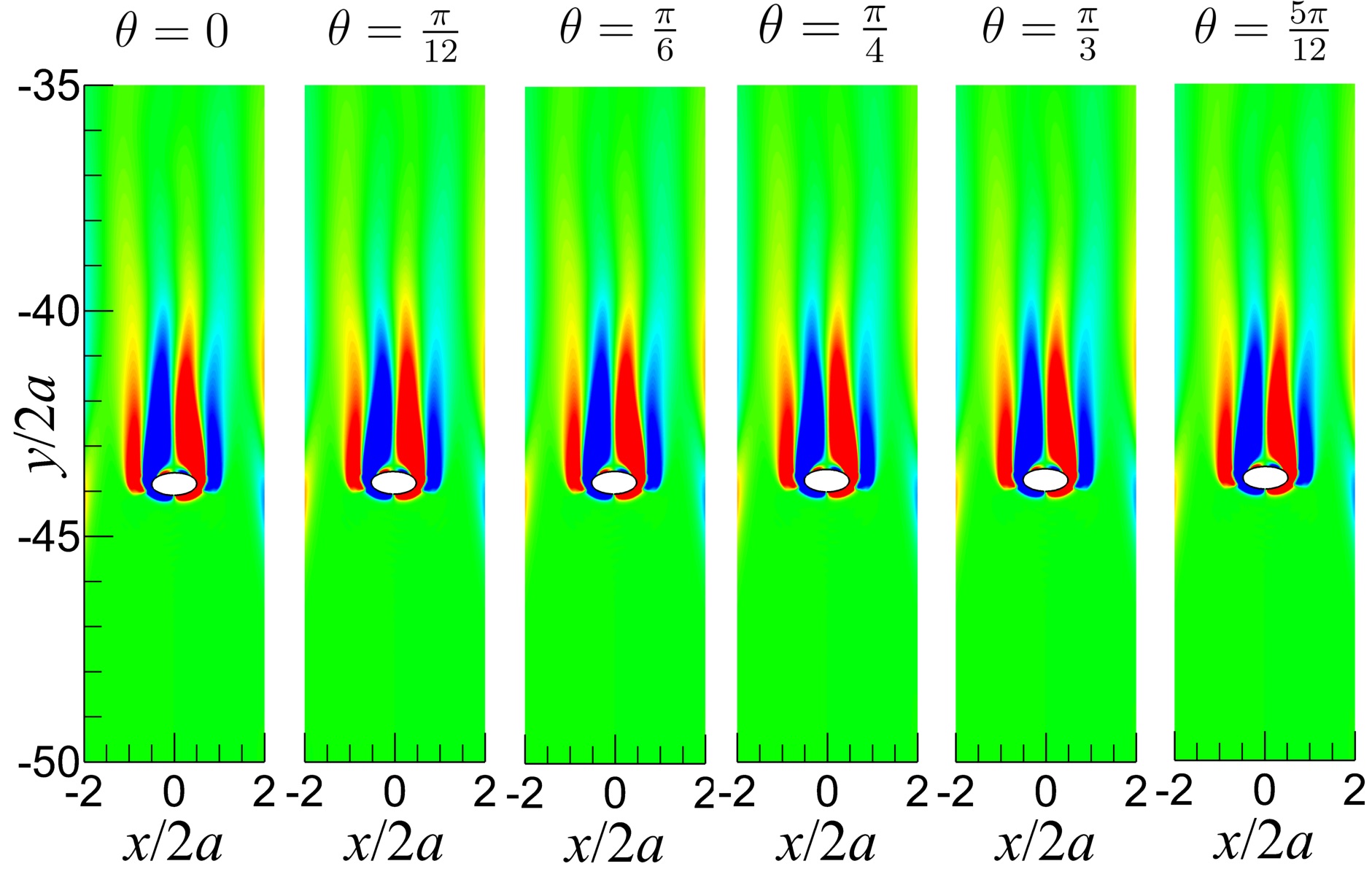}
\caption{Vorticity fields at $\bar{t}=1$ for different initial orientaton angles with $\rho^*=1.5$ and $N_\text{m}=3\text{e}4$.}
\label{fig:Sedi_vor_mag30_ini_phase}
\end{figure}

Anderson~et~al.~\cite{andersen2005unsteady} reported that the initial orientation angle will affect the dynamics of a sedimenting particle. Therefore, it is meaningful to study the effect of the initial orientation angle to the electromagnetic force control of the particle. Here, the density ratio is set to 1.5 and the strength of the electromagnetic force is chosen as $N_\text{m}=3\text{e}4$. We choose the initial orientation angle ranges in $[0,\frac{5\pi}{12}]$ with an interval of $\frac{\pi}{12}$. Fig.~\ref{fig:Sedi_vor_mag30_ini_phase} shows the vorticity field at $\bar{t}=1$ for different initial orientation angles. Under the control of the electromagnetic force, the unsteady wakes of the particle are controlled for all the cases and the sedimenting distances are very close.

%\FloatBarrier

\begin{figure}[htb]
\centering
\subfloat[]{\label{fig:Sedi_dis_mag30_ini_phase}\includegraphics[width=7.5cm]{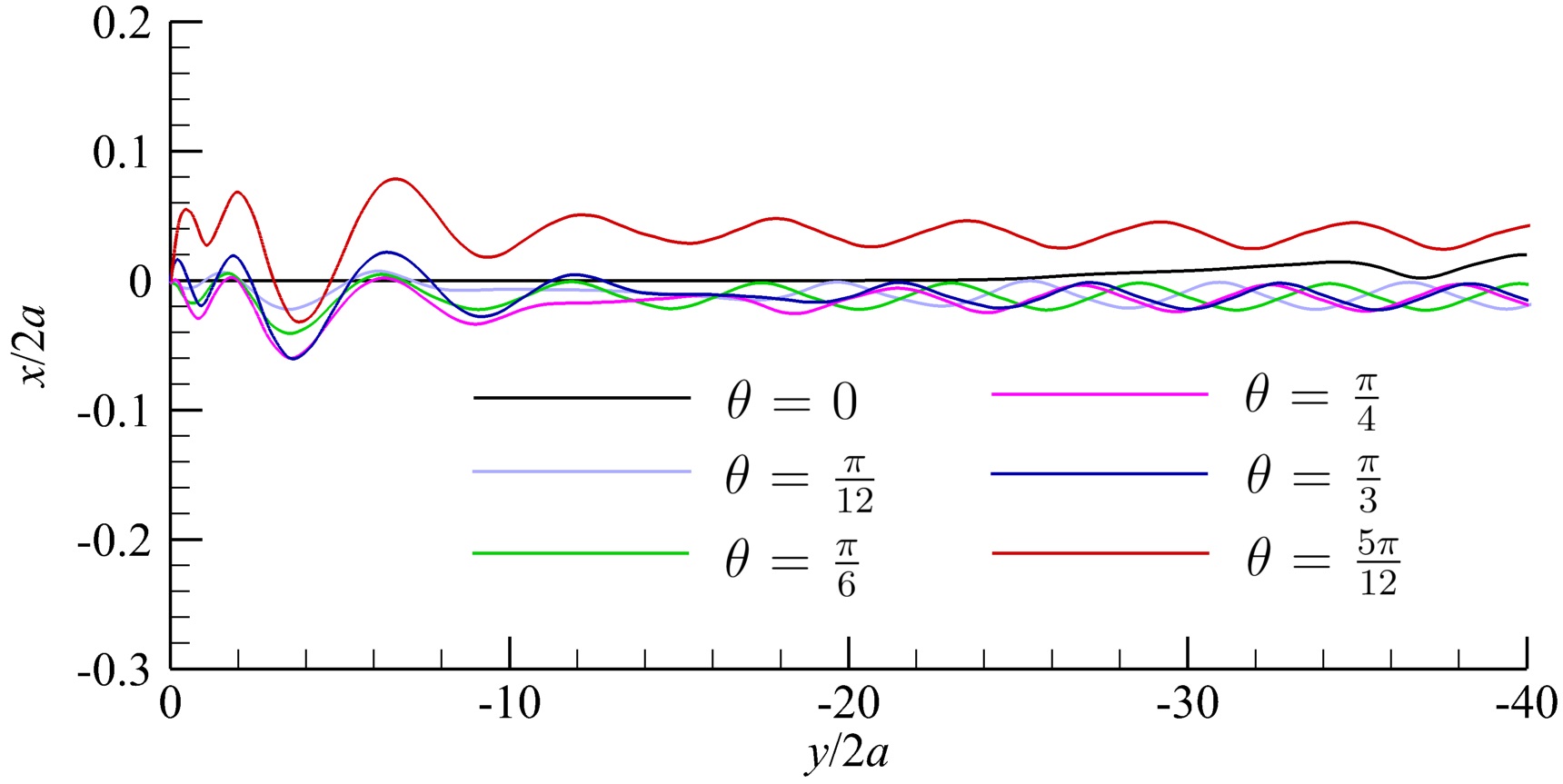}}\,\,
\subfloat[]{\label{fig:Sedi_ang_mag30_ini_phase}\includegraphics[width=7.5cm]{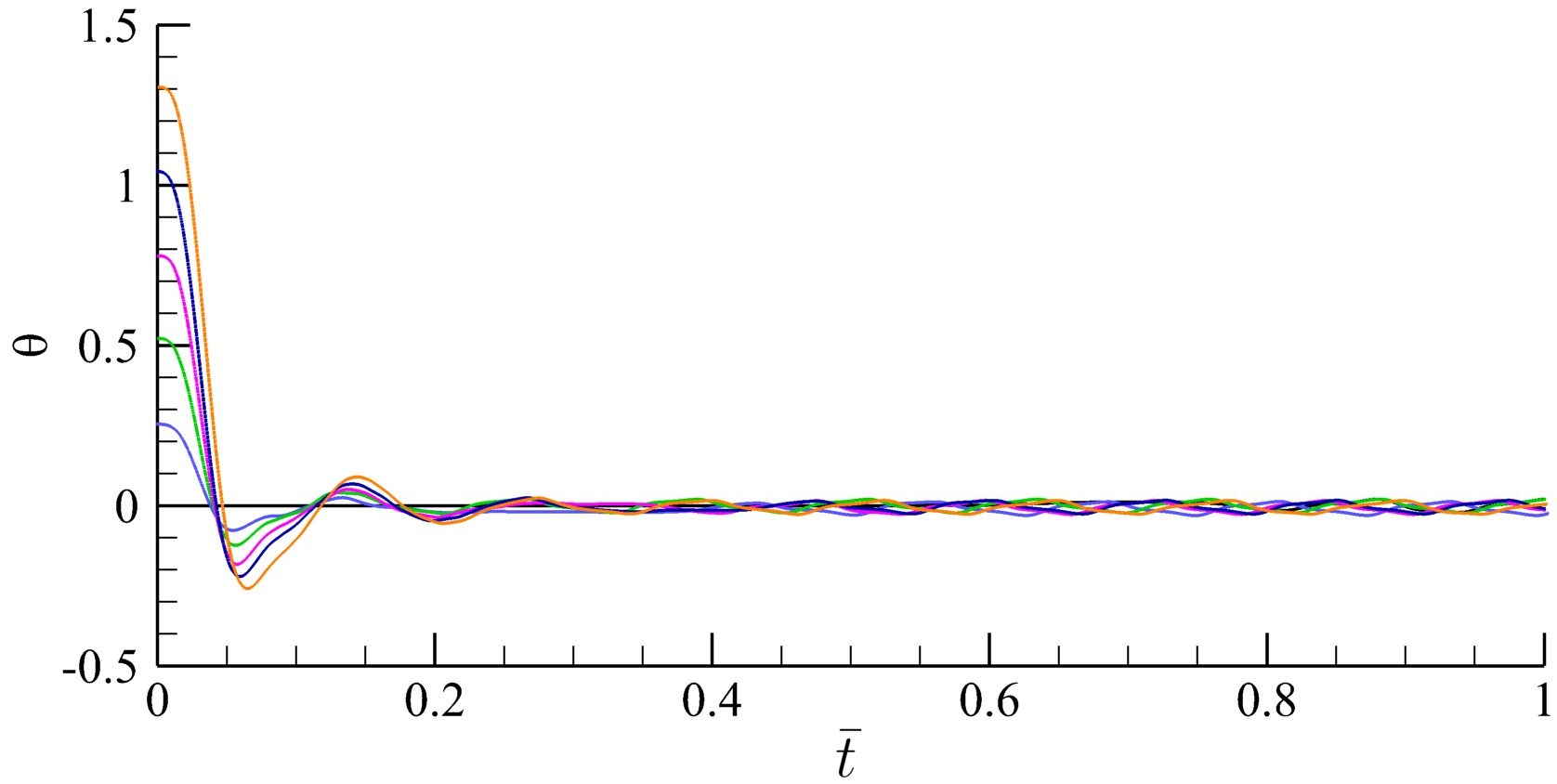}}
\caption{(a) Trajectories of particle center for different initial orientaton angles, and (b) Orientation angle \textcolor{black}{against the dimensionless time} for different initial orientation angles with $\rho^*=1.5$ and $N_\text{m}=3\text{e}4$.}
\label{fig:Sedi_dis_ang_ini_phase}
\end{figure}

The evolution of particle trajectories and orientation angles are shown in Fig.~\ref{fig:Sedi_dis_ang_ini_phase}\subref{fig:Sedi_dis_mag30_ini_phase} and Fig.~\ref{fig:Sedi_dis_ang_ini_phase}\subref{fig:Sedi_ang_mag30_ini_phase} to further understand the effect of the initial orientation angle to the electromagnetic force control of the particle sedimentation. For $\theta=0$, the particle almost moves in a straight line when $y/2a>-30$. Subsequently, the particle center deviates the center of the domain in the $x$ direction with a small deviation amplitude. For $\theta > 0$, including the nearly vertical case, the particle trajectories all show small oscillations around the $x$ direction ($|x/2a|<0.1$). Although the trajectories are slightly different for different initial orientation angles, the sedimentation processes are all much more stable than without adding the external electromagnetic force. This means the electromagnetic control approach is robust for various initial orientation angles. From Fig.~\ref{fig:Sedi_dis_ang_ini_phase}\subref{fig:Sedi_ang_mag30_ini_phase}, the rotational motions of the particles are all well controlled by using electromagnetic forces for different initial orientation angles. Hence the initial orientation angle has little effect on the electromagnetic force control of the sedimenting particle.

%\FloatBarrier

%\begin{figure}[htb]
%\centering
%\includegraphics[width=9cm]{Sedi_ang_mag30_ini_phase}
%\caption{Orientation angle \textcolor{black}{against the dimensionless time} for different initial orientation angles and $N_\text{m}=3\text{e}4$.}
%\label{fig:Sedi_ang_mag30_ini_phase}
%\end{figure}
%\FloatBarrier

\subsection{Effect of the density ratio}
\begin{figure}[htb]
\centering
\includegraphics[width=16cm]{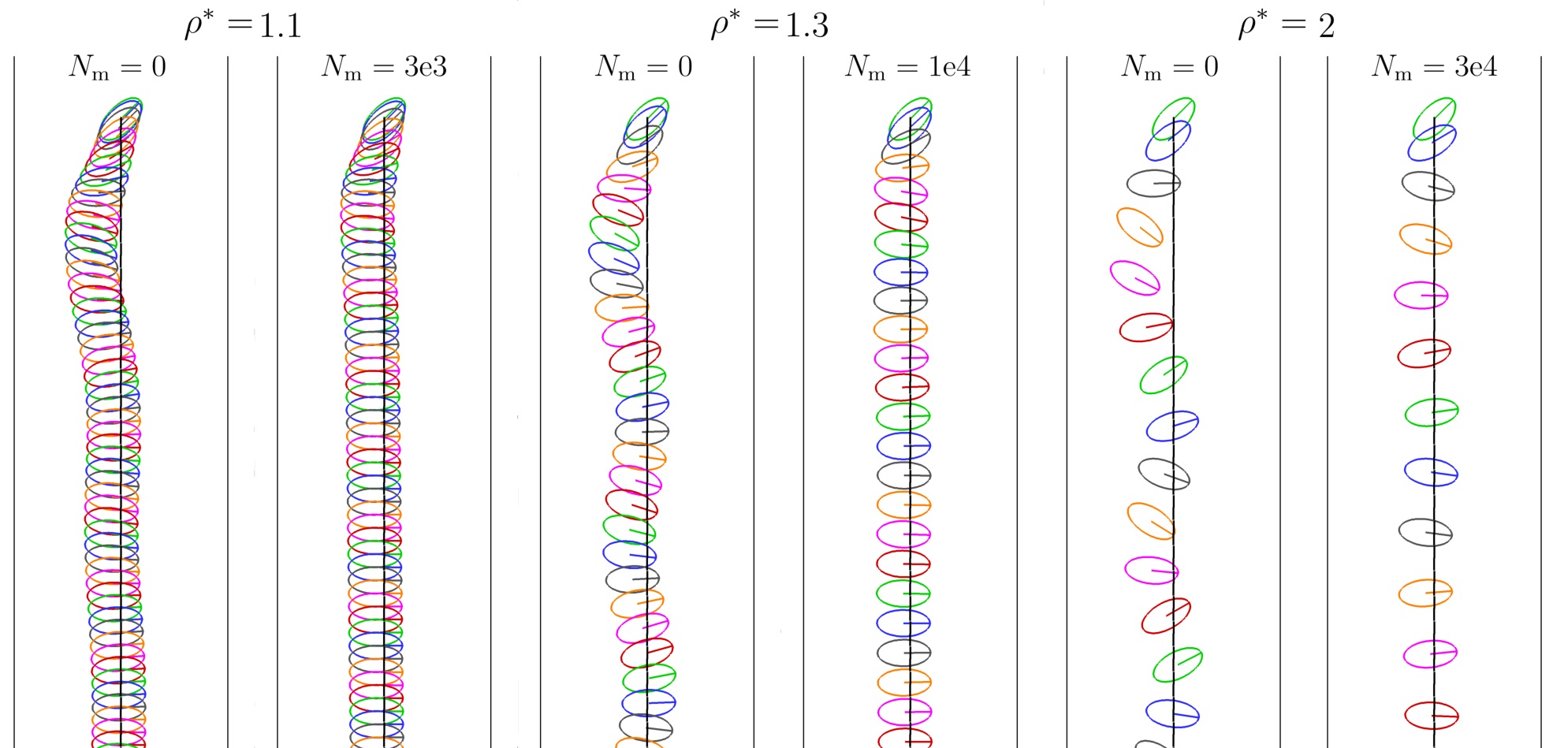}
\caption{\textcolor{black}{Sedimentation processes for the elliptical particles with and without electromagnetic forces for $\rho^*=1.1$, 1.3 and 2.}}
\label{fig:angle_dyn_11}
\end{figure}

\begin{figure}[htb]
\centering
\includegraphics[width=0.95cm,valign=t]{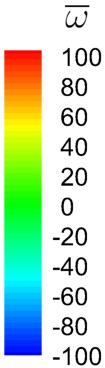}
\includegraphics[width=7cm,valign=t]{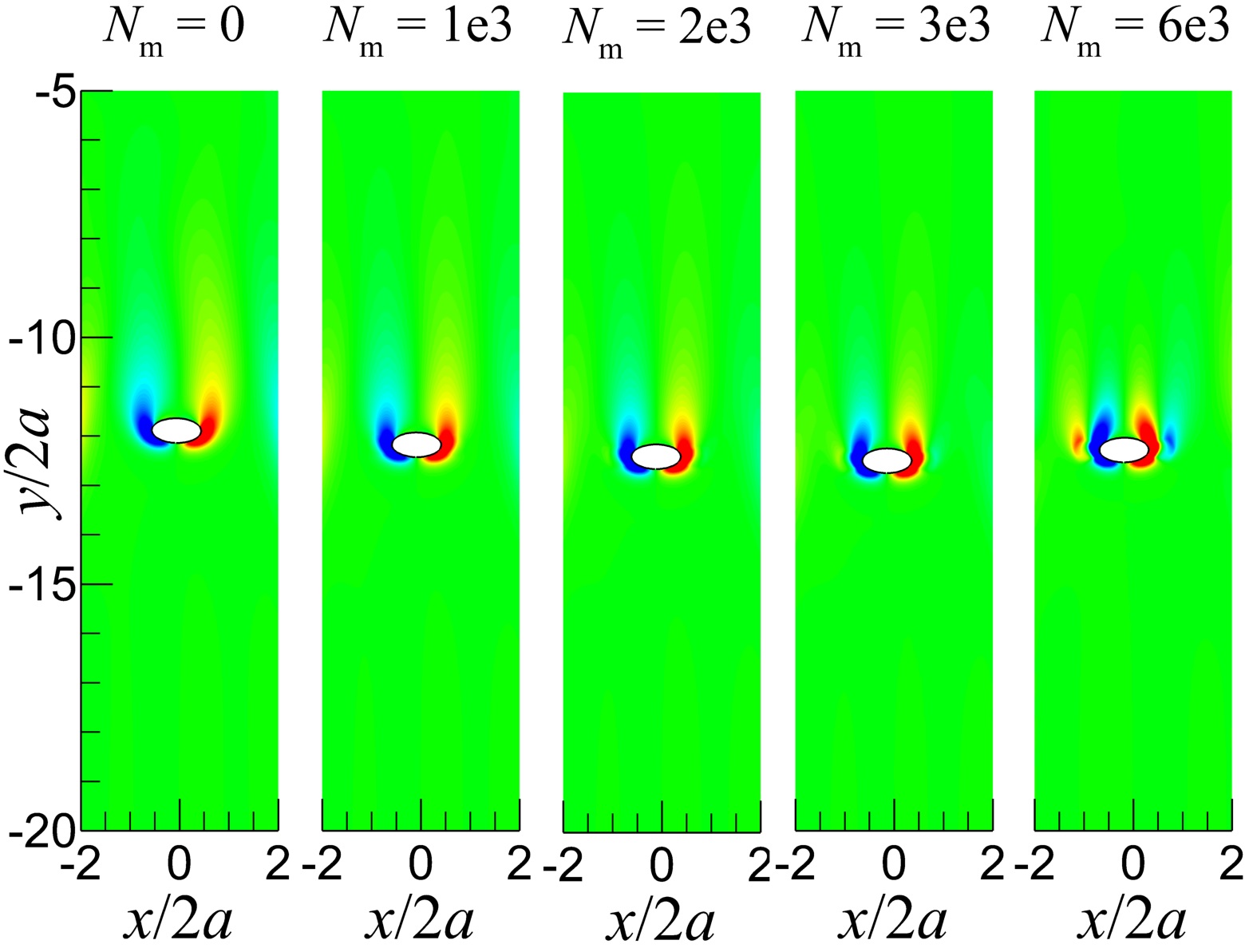}
\caption{Vorticity fields at $\bar{t}=1$ for different magnitudes of electromagnetic forces when $\rho^*=1.1$.}
\label{fig:Mag_vor_whole_density1.1}
\end{figure}

\textcolor{black}{The Reynolds number is determined by the density ratio when the viscosity of the fluid and the geometry of the particle remain unchanged. Three different density ratios of $\rho^*$= 1.1, 1.3 and 2 are chosen to check the robustness of the controlling approach. Fig.~\ref{fig:angle_dyn_11} presents the comparisons between the sedimentation processes for the three density ratios with and without the electromagnetic control. For $\rho^*$= 1.3 and 2.0 in the pure hydrodynamic case, periodic oscillations of the particle trajectories are similar to that of $\rho^*$=1.5. The electromagnetic force is shown to be effective in controlling the rotational motions of the particles in these two density ratios.} When the density ratio decreases to 1.1, vortex shedding will not appear even without the electromagnetic force and the motion of the elliptical particle tends to a steady descent~\cite{xia2009flow}. Fig.~\ref{fig:Mag_vor_whole_density1.1} shows the instaneous vorticity fields of $\rho^*=1.1$ under the control of different strengths of electromagnetic forces. Different from the $\rho^*=1.5$ cases in Fig.~\ref{fig:Mag_vor_whole}, the electromagnetic force will lower the sedimentation speed when the magnitude of the force is too strong. In addition, the vorticity induced by the electromagnetic force becomes important when $N_\text{m}=6\text{e}3$, and cause the particle to deviate the $x$ center of the domain. Therefore, the strength of the electromagnetic force needed for best control of the sedimentation process is one order of magnitude smaller for $\rho^*=1.1$ than for $\rho^*=1.5$.

%\FloatBarrier

\begin{figure}[htb]
\centering
\subfloat[]{\label{fig:25dx_dis_mag_5}\includegraphics[width=7.5cm]{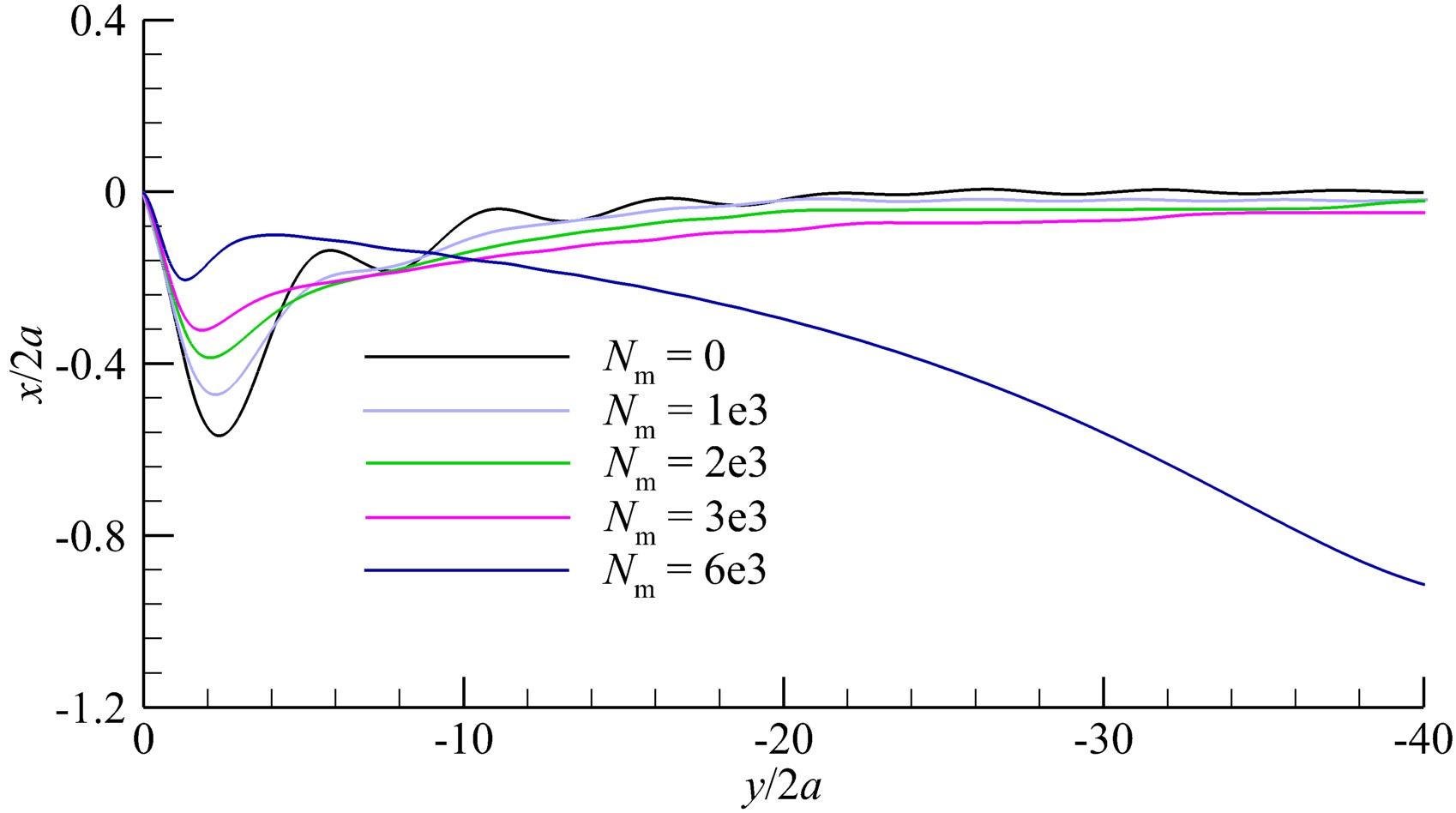}}\,\,
\subfloat[]{\label{fig:Sedi_mag_ang1point1}\includegraphics[width=7.5cm]{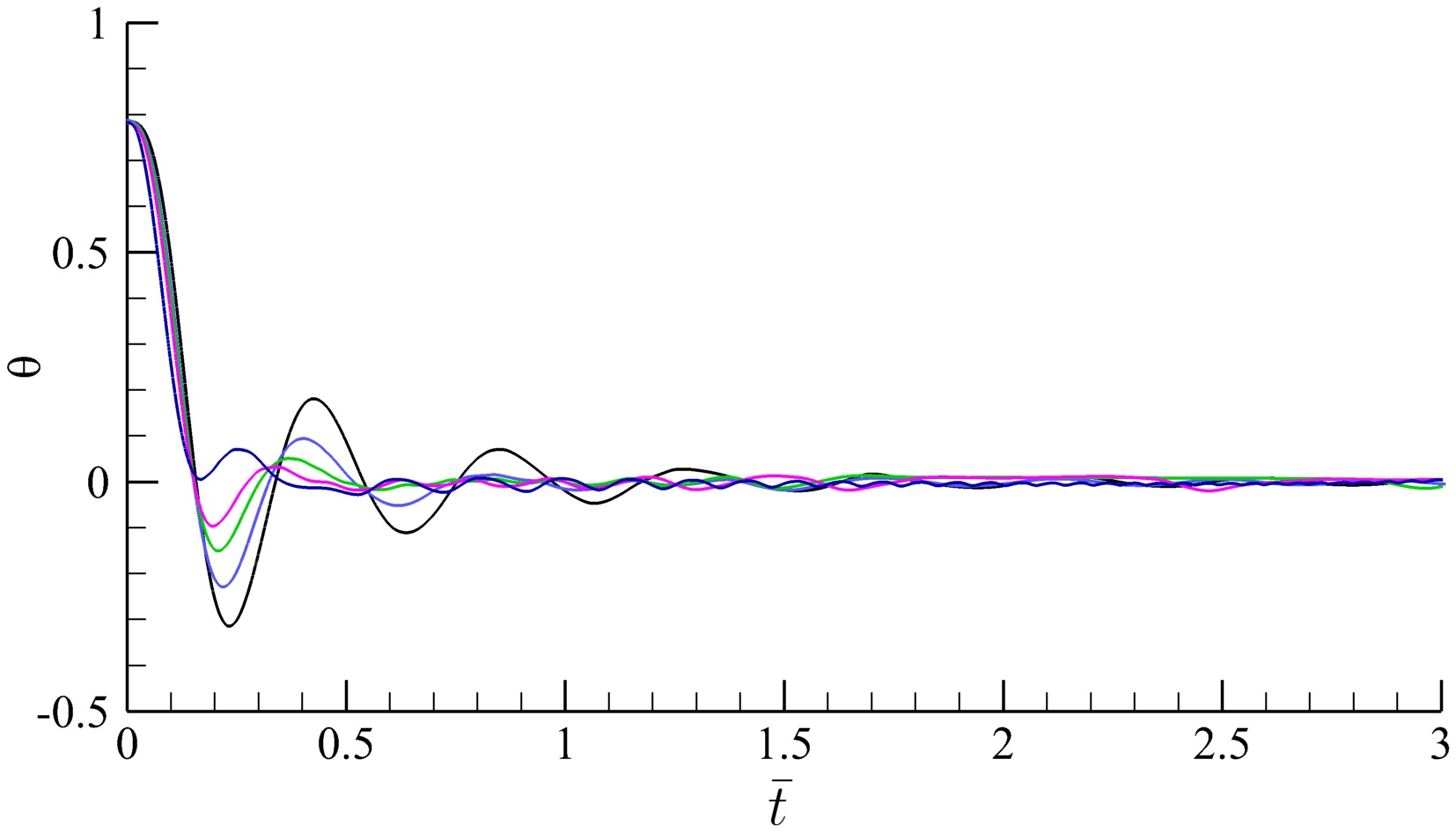}}
\caption{(a) Particle trajectories for different magnitudes of electromagnetic forces, and (b) orientation angle \textcolor{black}{against the dimensionless time} for different magnitudes of electromagnetic forces when $\rho^*=1.1$.}
\label{fig:dis_ang_1.1}
\end{figure}

Fig.~\ref{fig:dis_ang_1.1}\subref{fig:25dx_dis_mag_5} and Fig.~\ref{fig:dis_ang_1.1}\subref{fig:Sedi_mag_ang1point1} show the trajectories and the evolution of orientation angles of sedimenting particles for different magnitudes of electromagnetic forces in the case of $\rho^*=1.1$. When $0\leq N_\text{m}\leq 3\text{e}3$, the oscillation in the trajectories become smaller for stronger electromagnetic forces. This indicates that larger electromagnetic forces are more effective in controlling the sedimentation process. However, for the case of $N_\text{m}=6\text{e}3$, an obvious deviation in the $x$ direction appears because the electromagnetic force is able to balance the pressure differences on the left and right side of the particle even when the orientation angle is small. For this reason, to make the elliptical particle sediments stably, the electromagnetic force should also not be too excessive. It should be mentioned that the particle deviates towards the left wall instead of the right wall as shown when $\rho^*=1.5$. The difference is caused by the low density ratio so that the boundary layer forms slowly even for $N_\text{m}=6\text{e}3$ and the lift force is larger than the pressure differences on the left and right side of the ellipse. Again, the significant magnetic induced vorticity makes the particle moves towards the wall continuously. 
%\FloatBarrier

%\begin{figure}[htb]
%\centering
%\includegraphics[width=9cm]{Sedi_mag_ang1point1}
%\caption{Orientation angle \textcolor{black}{against the dimensionless time} for different magnitudes of electromagnetic forces when $\rho^*=1.1$.}
%\label{fig:Sedi_mag_ang1point1}
%\end{figure}
%\FloatBarrier

\begin{figure}[htb]
\centering
\includegraphics[width=9.5cm]{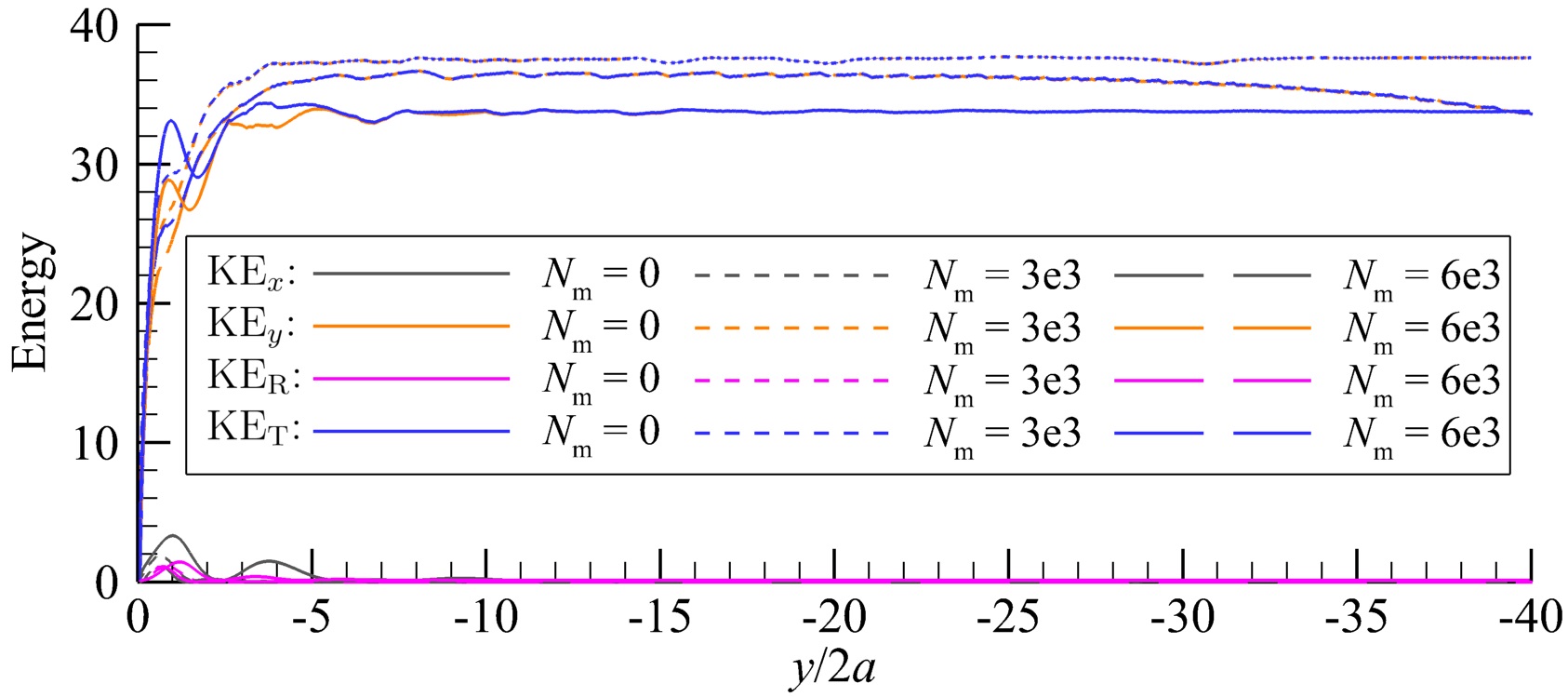}
\caption{Comparison of particle energies for $N_\text{m}=0$, $3\text{e}3$ and $6\text{e}3$ when $\rho^*=1.1$.}
\label{fig:energy1point1}
\end{figure}

Shown in Fig.~\ref{fig:energy1point1} are the energies of the particle during sedimentation for three different magnitudes of electromagnetic forces. After $y/2a=-6$, the total energy is almost equal to the kinetic energy in the $y$ direction for all the cases. Because of the initial orientation angle, the elliptical particle will rotate to generate additional vorticities. Due to the suppression of vorticity generation by the electromagnetic force, the total energy for the $N_\text{m}=3\text{e}3$ case is larger than the $N_\text{m}=0$ case. However, the total energy for the $N_\text{m}=6\text{e}3$ case is smaller than the $N_\text{m}=3\text{e}3$ case. This is different from Fig.~\ref{fig:Energy_comversion} when $\rho^*=1.5$, in which the total energy for the $N_\text{m}=1.5\text{e}4$ case is greater than the $N_\text{m}=0$ case. The loss of the total energy for $\rho^*=1.1$ with the excessive electromagnetic force is caused by that the particle energy is greatly influenced by the work done by the controlling force. The loss of the kinetic energy in the $y$ direction causes the particle to sediment slower for the $N_\text{m}=6\text{e}3$ case than the $N_\text{m}=3\text{e}3$ case (see Fig.~\ref{fig:Mag_vor_whole_density1.1}).
%\FloatBarrier

\section{Conclusions}
This paper has investigated the electromagnetic control of a sedimenting elliptical particle by using the immersed interface-lattice Boltzmann method. Two main mechanisms that contribute to the electromagnetic control of the particle motion are presented, i.e. suppressing vorticity generation on the boundary layer and reducing shedding shedding vortices. The former mechanism is demonstrated by the vorticity equation, meanwhile the later one is   explained by the streamwise momentum equation. We consider various strengths of electromagnetic forces to control a sedimenting particle of density ratio ($\rho^*$) equalling to 1.5. Without the electromagnetic force, vortex shedding behind the particle appears and the trajectory of the particle tends to be periodic. By adding electromagnetic forces, vortex shedding can be suppressed and the elliptical particles sediment more stably. Also, the results show that particle motions are affected by electromagnetic forces, including less rotations and smaller oscillation amplitudes in the $x$ direction. As the strength of the electromagnetic force increases, obvious magnetic induced vorticities appear. When the electromagnetic force is excessive, the particle center will deviate from the center of the computational domain in the $x$ direction. \textcolor{black}{Because less potential energy is converted into the rotational energy and the kinetic energy in the $x$ direction, the terminal velocity of the elliptical particle increased about 40\% with the control of the electromagnetic force.} Using $N_\text{m}=3\text{e}4$ and $\rho^*=1.5$, we change the initial orientation angles of the elliptical particle, in which $N_\text{m}$ is the dimensionless electromagnetic force strength. We show that the electromagnetic force can effectively control the sedimentation for various orientation angles. \textcolor{black}{In addition, the electromagnetic control approach is robust for different density ratio based on our limited calculations.}

When $\rho^*$ changes to 1.1, the electromagnetic force can decrease the oscillation amplitude of the particle in the $x$ direction. Like in the $\rho^*=1.5$ cases, the vorticity induced by the electromagnetic force will deviate from the particle mass center to the side walls once the electromagnetic force is excessive. Moreover, the electromagnetic force needed to get the best controlling performance is one order of magnitude less than that in the $\rho^*=1.5$ cases. Compared to the $\rho^*=1.5$ cases, one difference is that the particle sediments slower if the strength of the electromagnetic force is excessive. This is due to the direction of the electromagnetic force is opposite to the gravity direction and the total energy of the particle is decreased in case the electromagnetic force is excessive.

\section{Data availability}
The data that support the findings of this study are available from the corresponding author upon reasonable request.

\section{Acknowledgements}
This work is partially supported by NSFC Basic Science Center Program for ``Multiscale Problems in Nonlinear Mechanics'' (NO.~11988102). \textcolor{black}{The authors would like to thank Dr.~Xiaolei Yang for many discussions. The authors also thank the reviewers for the comments to improve this paper.}
%\FloatBarrier

{
\linespread{0.5} \selectfont
%\begingroup
%\setstretch{0.8}
%\printbibliography
\bibliographystyle{unsrt}
\bibliography{myref}
%\endgroup
}

\end{document}